\documentclass[graybox]{svmult}

\usepackage{type1cm}        % activate if the above 3 fonts are
\usepackage{makeidx}         % allows index generation
\usepackage{graphicx}        % standard LaTeX graphics tool
\usepackage{multicol}        % used for the two-column index
\usepackage[bottom]{footmisc}% places footnotes at page bottom
\usepackage{tikz}

\usepackage{tabularx}

\usepackage{newtxtext}       % 
\usepackage[varvw]{newtxmath}       % selects Times Roman as basic font

\makeindex             % used for the subject index
\begin{document}
%\tableofcontents % <------------------- to remove!!!

\title*{The benefits and costs of agglomeration: insights from economics and complexity}
\author{Andres Gomez-Lievano and Michail Fragkias}
%\institute{Andres Gomez-Lievano \at Health Economics and Outcomes Research, Analysis Group, Inc., 111 Huntington Ave, Boston, MA 02199, USA 
%\and Growth Lab, Harvard University, 79 John F. Kennedy St, Cambridge, MA 02138, USA 5\email{andres\_gomez@hks.harvard.edu}}
\institute{Andres Gomez-Lievano \at Analysis Group, Boston, MA, USA \\ 
Growth Lab, Harvard University, Cambridge, MA, USA \\ 
\email{andres\_gomez@hks.harvard.edu} \\
\\
Michail Fragkias \at Department of Economics, Boise State University, Boise, ID, USA \\ 
\email{michailfragkias@boisestate.edu}}


\maketitle

\abstract{There are many benefits and costs that come from people and firms clustering together in space. Agglomeration economies, in particular, are the manifestation of centripetal forces that make larger cities disproportionately more wealthy than smaller cities, pulling together individuals and firms in close physical proximity. Measuring agglomeration economies, however, is not easy, and the identification of its causes is still debated. Such association of productivity with size can arise from interactions that are facilitated by cities (``positive externalities''), but also from more productive individuals moving in and sorting into large cities (``self-sorting''). Under certain circumstances, even pure randomness can generate increasing returns to scale. In this chapter, we discuss some of the empirical observations, models, measurement challenges, and open question associated with the phenomenon of agglomeration economies. Furthermore, we discuss the implications of urban complexity theory, and in particular urban scaling, for the literature in agglomeration economies.}

%%%%%%%%%%%%%%%%%%%%%%%%%%%%%%%%%%%%%%%%%%%%%%%%%%%%%%
\section{Introduction}
\label{sec:chAGL:intro}

The study of \emph{agglomeration economies} explores the economic benefits that arise when individuals live and work \emph{in close physical proximity}. The topic originates in the observation of the existence of strong and persistent geographical concentrations in economic life: production and consumption activities in particular. The idea of agglomeration economies has been one of the most central topics studied by urban economists and economic geographers, if not \emph{the} most studied one \cite{glaeser2008cities}. 

The reaping of agglomeration economies has historically been identified as a central issue in our quest for increasing economic productivity and eventual prosperity at the local, regional and global scale. But shocks in the way we live and work may have a significant effect in the trajectory of those economies from spatial concentration. In some cases, we can ask whether these economies will continue to benefit humanity in the long term. For example, the COVID-19 pandemic has shown that households and firms are able to re-negotiating life/work arrangements, affecting the spatial patterns of living and working. Depending on the outcome of this process, and the economies present in complementary technologies for the exchange of ideas (e.g. remotely), we can expect important differentials in the level of agglomeration economies experienced and resulting economic prosperity.

There is a tendency to think that cities exist \emph{because} of agglomeration economies. The reasoning goes that since cities are ``the absence of space between people and companies'' (as the urban economist Edward L. Glaeser said \cite[p.6]{Glaeser2011Totc}), if there are benefits to agglomerating, people will tend to seek each other's company, and cities will naturally form. According to this reasoning, cities are the physical manifestation of the gains from being close to one another. However, there is a certain circularity here, leading in turn to a misconception: that one concept (cities) cannot exist without the other (agglomeration economies). That both are synonyms. But they are not. So to understand what agglomeration economies are, it is useful to clarify this difference.

On the one hand, cities can exist because of reasons not associated with any economic benefits from living and working in high densities. For example, people can come together inside some walls for religious reasons, to protect themselves against enemies or nature, or to be next to some natural resource like a river, forest, mine, or port \cite{sullivan2006firstcities}. On the other, there are many disadvantages from living in cities, like traffic, pollution, high rents, crime, and disease. Thus, agglomeration economies are only an aspect of cities. %it is not obvious why agglomerating would be beneficial. 

The benefits and costs from agglomerating act as two opposing forces. The former as centripetal forces that pull people (and companies) together, and the later as centrifugal forces that push them away.
In recent times, cities have become less dependent on being close to certain natural resources. At the same time, evidence tell us that the benefits from agglomeration are dominating its costs, and this has been reflected in the fast pace of urbanization and the continued growth of cities that we see today around the world. Hence, agglomeration can be considered one of the most important forces of our current world, and we better understand its effects (i.e., agglomeration economies). 

In what follows, we address several dimensions of agglomeration economies and several open research questions in this area. Section \ref{sec:chAGL:empirics} reviews the basic and advanced notions surrounding agglomeration economies - how they have been theoretically conceptualized, empirically measured, and mathematically modeled; the challenge of causal inference and differing experiences across parts of the world. Section \ref{sec:chAGL:smodel} discusses ideas from economics and complexity science that provide a more nuanced understanding of agglomeration phenomena: the notion of spatial equilibrium from economics (in contrast to the idea of disequilibrium emerging complexity science); the microfoundations of agglomeration economies; artificial increasing returns to scale; and urban scaling theory. Section \ref{sec:chAGL:complexity} explores the possibility of blending mainstream economic analysis of agglomeration with the complexity science perspective. Section \ref{sec:chAGL:beyond} reviews agglomeration in the context of sociology, cultural and evolutionary anthropology, pointing to the possibility of a grand synthesis regarding agglomeration phenomena across fields. Finally, section \ref{sec:chAGL:conclusions} concludes and discusses open questions that still remain.

%%%%%%%%%%%%%%%%%%%%%%%%%%%%%%%%%%%%%%%%%%%%%%%%%%%%%%
\section{The empirics of agglomeration economies}
\label{sec:chAGL:empirics}

\subsection{Conceptual and technical frameworks}
Agglomeration economies, broadly construed, are the benefits that emerge from the clustering of (economic) activity closely in space. In order to quantify and estimate them, we will use a more precise definition: \emph{agglomeration economies} are the \emph{increasing returns to scale} (IRS), as well as \emph{external economies specific to a location}, related to economic productivity and well-being, occurring in specific locations, such as cities, as they get larger and/or denser. In the field of economic geography, increasing returns to scale play a significant role in the location choice of firms, leading to the formation of core-periphery patterns  \cite{krugman1991increasingreturns}.

Agglomeration in economics has been a central theme of study in urban and regional economics and the new economic geography \cite{fujita1999spatial}. Specifically in economics, agglomeration economies are economies that are external to firms, but internal to a location \cite{goldstein1984economies}. Four types of agglomeration economies based on scale are conceptualized in this framework: (a) economies internal to the firm at a given location - that is economies from mass production, (b) economies external to a firm at a specific location, (c) localization economies (internal to an industry at that location) and (d) urbanization economies (external to both firm and industry at a particular location).

The ``returns to scale'' of a quantity $F$ (e.g., the total well-being in a settlement) with respect to a variable $x$ (e.g., the size of the settlement) is given by the change in the function when $x$ is multiplied a constant $\lambda>1$.\footnote{Since the returns to scale concept in economics originates in production theory, we utilize this concept in line with a ``city as a firm'' analogy \cite{fujita1999spatial}. That is, in this exposition of the concept, we can think of a city as represented by a single firm.} Increasing returns occur when $F(\lambda x) > \lambda F(x)$. Let $f(x)=F(x)/x$ be the \emph{per capita} value. Then, it is easy to verify that IRS implies that $f(\lambda x) > f(x)$. The quantity $f$ can be wages, patenting rates, or other measures of individual productivity in a given city, and the quantity $x$ is the total employment, the working age population, the total population size or even can be the population density of the city. The discussion of increasing returns to scale leads into the topic of economies of scale when the payment of factors of production is introduced.

Economists perceive external economies of agglomeration as resulting from a shift in a production function of a firm due to the effects of a change in scale of the area that a firm exists in. In the simplest form, this concept can assume $Y = G(N)F(K, L)$, where $N$ is population size, $K$ is capital and $L$ is labor. Expressed in terms of output per worker - dividing by $L$, we get $y = G(N)f(k)$.\footnote{Here, we are assuming the function $F(K,L)$ to have \emph{constant} returns to scale. That is, $F(\lambda K, \lambda L)=\lambda F(K,L)$ for any positive constant $\lambda$. In particular, if $\lambda=1/L$, we get that $(1/L)F(K,L)=F(K/L, 1)\equiv f(k)$, a function of capital per worker $k=K/L$.} The function $G(N)$ is referred to as a `Hicks-neutral' externality since we make the assumption that it does not affect the optimal choice of labor and capital in the production process.\footnote{A Hicks-neutral externality raises the marginal product of labor and capital proportionally at a specific capital to labor ratio.} Formally, if $dG/dN>0$, we observe economies of agglomeration while if $dG/dN<0$, diseconomies of agglomeration. Figure \ref{fig:prodfunct} shows economies of agglomeration by plotting a hypothetical production function of a firm existing in a larger agglomeration ($N'$) vs a smaller agglomeration ($N$). Sveikauskas (1975) \cite{Sveikauskas1975} was among the first to explore the empirical relationship between city size and labor productivity from this point of view. His paper concluded that average wages were disproportionately higher in large cities because Hicks-neutral productivity was substantially higher in large cities.

\begin{center}
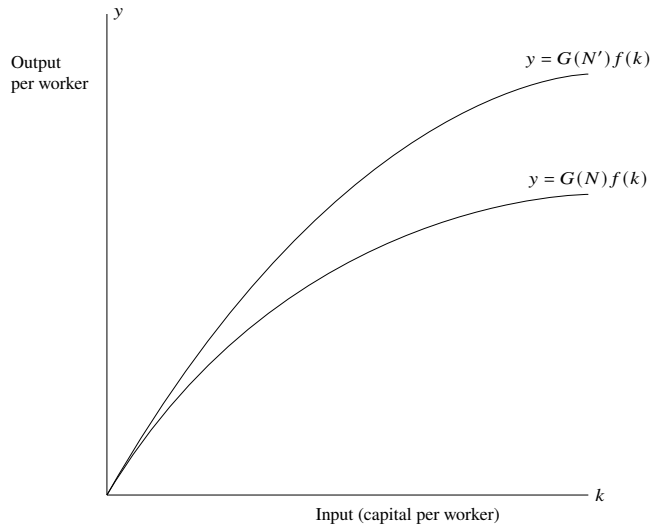
\begin{figure}
\scalebox{0.8}{
\begin{tikzpicture}(10,0)
\draw (0,0)--(8,0) node[right]{$k$}; 
\node [below] at (5,-0.1) {Input (capital per worker)};
\draw(0,0)--(0,8) node[right]{$y$};
\node [align=left, left] at (-0.2,7) {Output\\ per worker};
\draw(0,0) ..controls (3,5) and (8,5) .. (8,5) node[above]{$y=G(N)f(k)$};
\draw(0,0) ..controls (4,7) and (8,7) .. (8,7) node[above]{$y=G(N')f(k)$};
\end{tikzpicture}
}
\caption{External economies of agglomeration as experience by a firm through a shift in the production function due to an increase in the size of a city ($N$ to $N'$).}
\label{fig:prodfunct}
\end{figure}
\end{center}

\begin{figure}[t]
\centering
%\sidecaption[t]
\includegraphics[scale=.85]{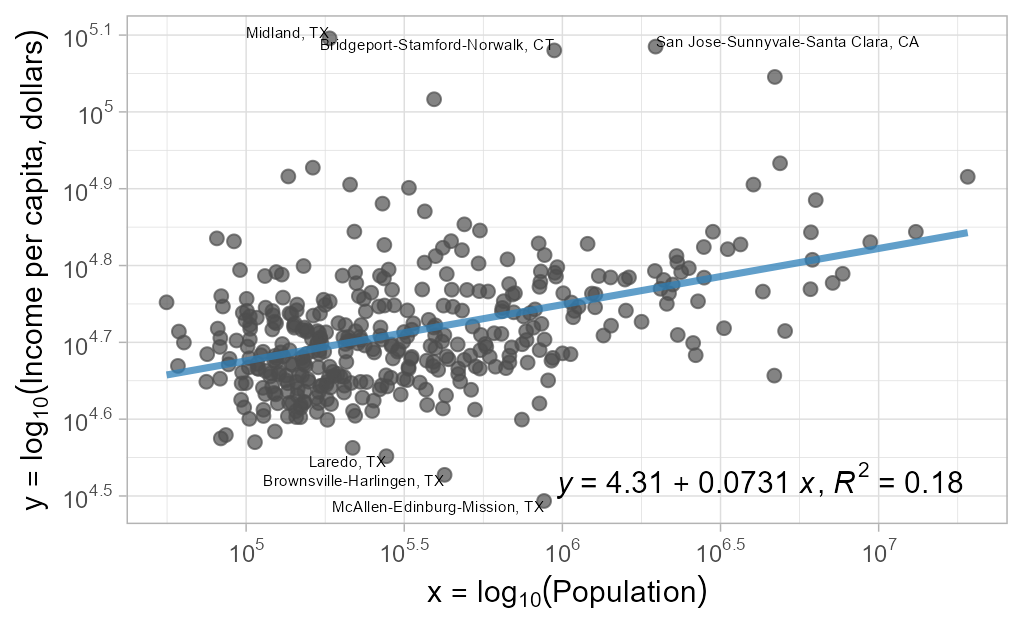}
\caption{Association between income per capita and population size across US MSAs in 2020. The name of the MSAs with the highest and lowest personal income are indicated in the figure. All dollar estimates are of current dollars (not adjusted for inflation). Source: Bureau of Economic Analysis (BEA) \cite{BEApersonalincomedata}.}
\label{fig:GMPvsN}       % Give a unique label
\end{figure}

Before going into models of agglomeration economies, the statistical technicalities, or the underlying causes, let us look at some data so that we understand why there is so much interest from urban economists about this phenomenon. As an example, let us start with data from Metropolitan Statistical Areas (MSAs)\footnote{MSAs are the most common unit of analysis in urban studies, and the US methodology for how to define them has been adopted by many other countries. Their delineation is not based on administrative or political boundaries (like those of counties, municipalities, provinces, states, countries, etc.), but on commuting patterns. The idea behind using commuting patterns to define urban boundaries is that cities are the places where people \emph{live and work}. MSAs account for approximately $86\%$ of the total US population.} in the United States (US) for the year 2020 in Figure~\ref{fig:GMPvsN}. There are 384 dots in the plot, each dot representing an MSA, and shows how the logarithm of average personal income correlates with the logarithm of population size. The use of a log-log plot to reveal agglomeration economies is not accidental. This is because the ``returns to scale'' of a function $f$ can be quantified by the ratio between the rate of increase of $f$ (defined as $\frac{1}{f}\partial f/\partial x$) and the rate of increase in $x$ (defined as $\frac{1}{x}\partial x/\partial x$). This is equivalent to comparing changes between logarithms, as $\partial \ln(f)/\partial \ln(x)$. As a consequence, a typical step into assessing the presence of agglomeration economies is plotting how $\ln(f)$ relates to $\ln(x)$. 

As can be seen in Fig.~\ref{fig:GMPvsN}, per capita income tends to be higher in larger cities. The coefficient of the regression is approximately $0.073$. Coefficients between logarithmic variables denote ``elasticities''. An elasticity of $0.073$ implies that a $1\%$ change in population size is approximately associated with a $0.073\%$ change in personal income.\footnote{The actual formula is that an $\epsilon$ percent change in the independent variable (e.g., city size) leads to a $[(1+\epsilon/100)^\beta -1]\times100$ percent change in the dependent variable (e.g., income). When $\epsilon$ is small, then $(1+\epsilon)^\beta -1\approx \beta \epsilon$, which means that a ``$\epsilon$ percent change in the independent variable corresponds to a $\beta \epsilon$ percent change in the dependent variable'', as was done above in the text. Sometimes, however, researchers interpret the elasticities with a $100\%$ change in the independent variable (e.g., doubling city size). In that case the ``$\beta \epsilon$'' formula will overestimate the effect on the dependent variable. With an elasticity of $7.3\%$, doubling the size of a city corresponds to income increasing by $5.2\%$.} 

It is worth noting that the relationship in Fig.~\ref{fig:GMPvsN} between log-population and log-income seems to be linear, and that the $7.3\%$ elasticity seems to apply across all city sizes. Such linear relationships in the form of $\ln(y)=A+B\ln(x)$, where the coefficient $B$ is a constant, describe a \emph{power-law} relationship between quantities $x$ and $y$, of the form $y=a x^B$ (where $a=e^A$)\footnote{If the relationship uses $\log_{10}$ instead of natural logarithms, then $a=10^A$.}. At the level of cities, agglomeration economies are often (though not always) represented with these power-law relationships. Mathematically, we therefore talk of evidence of agglomeration economies at the aggregate level when
\begin{equation}
    y=c N^\delta, 
\label{eq:percapitascaling}
\end{equation}
where $c$ is a constant, $N$ is city size, and, crucially, $\delta>0$ for variables $y$ that measure of economic well-being at a per-capita level.

%%%%%%%%%%%%%%%%%%%%%%%%%%%%%%%%%%%%%%%%%%%%%%%%%%%%%%%%%%%%%%%%%%%%
\subsection{Measurement issues: productivity and scale}
Going back to Fig.~\ref{fig:GMPvsN}, the power-law relationship between city size and personal income suggests the presence of agglomeration economies in US MSAs. Personal income, however, includes earnings not only from job salaries, but also from interests, dividends, transfers, and others sources\footnote{See \url{https://www.bea.gov/resources/learning-center/what-to-know-income-saving} (last accessed: May 15, 2022).}, which do not necessarily reflect the benefits accrued specifically due to living at a given location. For example, an individual might live in San Francisco but their income may come from renting an apartment in New York. This income would still be counted in San Francisco, even though the economic activity generating such income is from New York. So how can we measure the ``direct'' benefits from living in a particular city? A better quantity would be a measure of \emph{productivity}.

In economics, \emph{productivity} is an abstract measure of the performance in a production process. It has to do with the efficiency by which an agent (person, firm, industry, city, region, or country) converts inputs into output. Quantifying productivity is challenging because one must account for increases in output that are not trivial. For instance, if doubling the inputs of a particular production process ends up doubling the output produced, then the productivity of that process has not changed. This suggests productivity should be quantified as a ratio of $\frac{\text{output}}{\text{inputs}}$. At times, $\frac{\text{sales}}{\text{worker}}$ act as a proxy of productivity. But it is not that simple. There are more subtle effects to think about. For example, rather than considering doubling the inputs, consider substituting some type of inputs with another type (e.g., substituting workers with machines). A substitution may or may not induce a change in productivity. Which substitutions would suggest an increase in productivity? Should productivity be measured as $\frac{\text{profits}}{\text{costs of inputs}}$? 

Figure~\ref{fig:GMPvsN} also raises the question of whether total population size $N$ is the right measure of agglomeration. Would it be better to use \textit{population density} (that is, population per unit area)? Duranton and Puga (2020) \cite{duranton2020economics} argue that neither is a great choice: ``For instance, a highly concentrated but tiny cluster of economic activity is unlikely to generate strong agglomeration economies. On the other hand, workers located at the edge of large metropolitan areas are unlikely to benefit from their full scale in the job-matching process'' (p. 7). They also suggest that theory is not going to resolve this issue. Currently, the main practice in economics is to utilize total population, but increasingly, with the availability of better data, local density is also utilized

Thus, the economics of density literature moves away from exploring agglomeration economies as the relationship that emerges between measures of productivity and city size, and focuses on the alternative approach exploring the relationship of productivity against \textit{density}. Duranton and Puga (2020) show that urban density and population are highly correlated, assuming non-naive measures of urban density (e.g. that do not include extensive rural areas in the calculation). That is, proper density measures should reflect how density is experienced by the typical urban resident of the MSA.

The answers to the choice of measurement are not easy and change depending on the specific production process under consideration, and many other assumptions. We will not delve in such discussions here.\footnote{There is a whole field for how to measure productivity. See, e.g., \cite{felipe2003aggregation}. Similarly, for measuring density.} Instead, we will use \emph{nominal wages}, as is customarily done in the literature, to reflect the productivity of a place. (We will review the conceptual arguments in Section~\ref{sec:chAGL:puzzle}). We will mainly adopt scale as a measure of spatial concentration. As we will show, these choices allow us to discuss the literature from urban economics and urban scaling uniformly (see, however, \cite{ribeiro2018unveiling} and their use of density in the context of the urban scaling literature). Given these modeling choices, we can ask: are wages higher in larger cities?

%%%%%%%%%%%%%%%%%%%%%%%%%%%%%%%%%%
\subsection{Empirical estimation}
\label{sec:chAGL:estimation}

Let us define approximate productivity by taking a measure of total aggregate output generated by a single metropolitan area \textit{(i}) during a given year (e.g., total annual wage bill), $Y_i$ and dividing it by the number of workers in the city, $N_i$,
\begin{equation}
y_i=Y_i/N_i
\label{eq:YoverN}
\end{equation}
The ratio in Eq.~(\ref{eq:YoverN}) can be interpreted as the average productivity of workers in the city.\footnote{In general, when you take any measure of total output (e.g., total number of patents, total number cars produced, total value of good exported) and you divide it by the labor used to produce it, you get a measure of \emph{labor} productivity. As mentioned in the previous section, caution needs to be taken in the interpretation.}

Given data for $y_i$ and $N_i$ across several cities within a same urban system (like a country), the relationship between productivity and scale can be assessed using ordinary least squares (OLS) regression analysis. From the economics perspective, the mathematical model specification is derived by a competitive firm-level profit maximization problem solution \cite{moomaw1981productivity}. The statistical specification usually takes the form of a log-log model, given that researchers are exploring the relationship in terms of elasticities or estimating a power law (like in Fig.~\ref{fig:GMPvsN}). 

\begin{equation}
\ln(y_i)= \ln(c) + \delta \ln(N_i) + \epsilon_i, 
\label{eq:estimation_productivity_pop}
\end{equation}
where $\delta$ is the elasticity, and $\ln(c)$ is interpreted as the constant in standard regression. Some debate exists on whether ordinary least squares is the appropriate methodology in estimating power-laws, or whether maximum likelihood estimation or Bayesian econometrics should be utilized (see, e.g., \cite{shalizi2011scaling,GomezLievanoYounBettencourt2012,leitao2016scaling}).

Alternatively, given the land area of cities ($A_i$), one can also estimate the same model but replace the size of the city with a population density measure, $\rho_i = \frac{N_i}{A_i}$.

\begin{equation}
\ln(y_i)= \ln(c) + \delta \ln(\rho_i) + \epsilon_i, 
\label{eq:estimation_productivity_density}
\end{equation}

\begin{figure}[t]
%\sidecaption[t] 
\centering
\includegraphics[scale=.85]{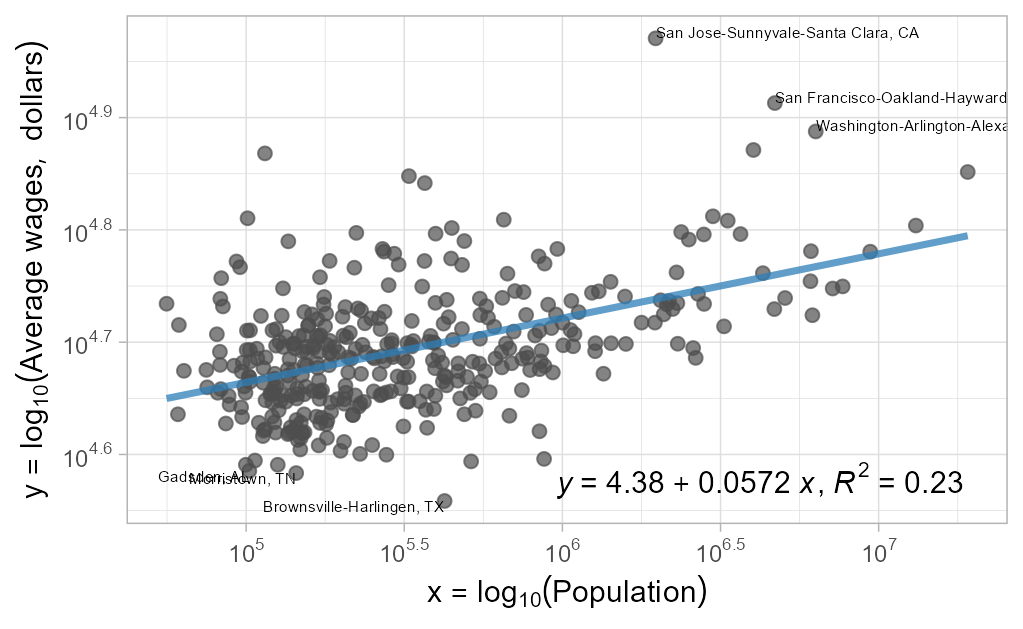}
\caption{Association between average wages and population size across US MSAs in 2020. The name of the MSAs with the highest and lowest average wages are indicated in the figure. Source: Occupational Employment and Wage Statistics (OEWS), Bureau of Labor Statistics (BLS) \cite{BLSwages2020}.}
\label{fig:wagesVSn}       % Give a unique label
\end{figure}

Typical results are shown in Figure~\ref{fig:wagesVSn}. Our results show a positive association between population size and well-being as measured by average wage. In this case, the estimated elasticity is $0.057$, implying that doubling the size of a city in this cross-section is associated with an increase of $4.0\%$ in average wages. This elasticity is often referred to as the \emph{urban wage premium}.

Similar results are found in the economics of density literature. Considering the agglomeration benefits when concentration is measured in terms of density, the literature provides similar results: Ahlfeldt and Pietrostefani \cite{AHLFELDT201993}, in a meta-analysis, measure the elasticity of productivity with respect to density at 0.04.

Estimates from simple bivariate regressions like Eqs.~(\ref{eq:estimation_productivity_pop}) and (\ref{eq:estimation_productivity_density}) have many limitations. They do not actually identify the \emph{causal} effect of what being close to other people does to productivity and well-being. They are aggregate correlations and, as such, they do not rule out other alternative causes. For example, the same correlation could emerge if people who are more productive prefer to be close to each other, and as a consequence, tend to move and \emph{self-sort} into (large) cities. As we mentioned in the introduction, another cause of both agglomeration (i.e., population growth) and higher productivity would be the presence of some natural advantages (potentially unobserved). 

With the above in mind, next, we review the empirical challenges of accurately measuring agglomeration economies.

%%%%%%%%%%%%%%%%%%%%%%%%%%%%%%%%%%
\subsection{Empirical challenges: causal inference}
\label{sec:chAGL:challenges}

\begin{figure}[t]
%\sidecaption[t] 
\centering
\includegraphics[scale=.55]{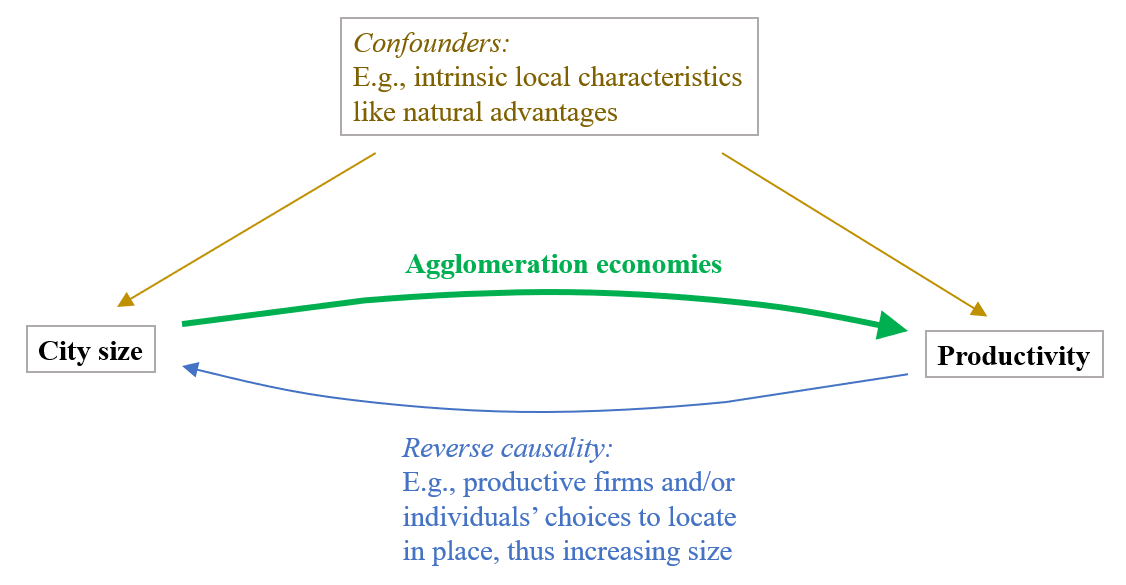}
\caption{Diagram of causes and effects that could explain correlations between city size and wages.}
\label{fig:causaldiagram}       % Give a unique label
\end{figure}

Quantifying agglomeration economies consists of being able to single out the \emph{causal effect} that physical proximity has on productivity. We have discussed this, concretely, in the case of the effect city size has on wages. The challenges involved can be illustrated in the causal diagram depicted in Figure~\ref{fig:causaldiagram}. Each box in the figure represents a variable, and each arrow represents a causal effect. Thus, measuring agglomeration economies amounts to quantifying the strength of the green arrow.

The main challenge in measuring the green arrow is that other arrows exist that also induce correlations between city size and productivity. These issues are represented by the top and bottom boxes in Fig.~\ref{fig:causaldiagram}. On the one hand, the top box and its yellow arrows represent variables that simultaneously cause cities to get larger \emph{and} firms to have higher productivities. An example is a local natural advantage, such as a river or a sea port, which may attract firms to operate there (causing city size to increase), and may be an important source of resources for the city (causing rises in productivity). These variables are called ``confounders''. They are a problem if they are unobserved. In general, however, the problem of confounders can be minimized by including available variables for such local urban characteristics as ``controls'' in regressions \cite{combes2010estimating}. For example, one could modify Eq.~(\ref{eq:estimation_productivity_pop}) to
\begin{equation}
\ln(y_i)= \ln(c) + \delta \ln(N_i) + \phi X_i+ \epsilon_i, 
\label{eq:estimation_productivity_pop_counfounders}
\end{equation}
where $X_i$ can be variables that stand for local characteristics of the city.

The real challenge comes from the bottom part of the diagram. That is, the blue arrow in Fig.~\ref{fig:causaldiagram} in which the direction of causality runs in the opposite direction as the one associated with the concept of agglomeration economies (green arrow). The fact that the arrow points in the opposite duration is not meant to suggest that there is something that counteracts or reduces the effect of agglomeration economies. Instead, the arrows just represent which variable are affected by changes in other variables. In particular, the green and blue arrows illustrate the fact that a positive association between city size and productivity might have two possible explanations: either city size causes higher productivity or individual cities that are already more productive tend to systematically grow to large sizes. The bottom part of the diagram represents the issue of so-called ``reverse causality'', in which economic agents  decide to locate themselves close to other agents that are inherently more productive, thus increasing the size of cities \cite{glaeser2001cities,Combes2011identification}. As a consequence, choices will create a correlation between size and productivity. According to this direction of causality, the correlation will emerge not because larger sizes cause economic activities to be more productive, but the other way around: city size will be the \textit{consequence} of the choices that productive individuals, firms and industries make. If high-productivity firms (or industries) tend to cluster in cities (increasing their size), the urban wage premium found in large cities could instead reflect firms' specific production processes rather than any benefits derived from geographical proximity. 

The issue of reverse causality illustrated in Fig.~\ref{fig:causaldiagram} is sometimes also referred to as ``self-sorting'', and it turns out to be incredibly hard to adjust for in statistical analyses that seek to measure the causal effect of agglomeration economies. In other words, one step in measuring the presence and strength of the green arrow is to ``shut down'' (statistically speaking) the effect of the blue arrow. However, difficulties include the fact that the production function of firms is often hard to observe, as well as the fact that it is hard to identify the reasons why firms and people migrate from city to city.

A full review of the details involved in separating the causal effect of agglomeration economies from other causal sources like confounding and self-sorting can be found in Combes and Gobillon (2015) \cite{CombesGobillon2015handbook}. The literature in urban economics keeps investigating new methodologies to address these empirical challenges, and the sophistication of the methods keeps increasing steadily, but most of the solutions have converged on a specific strategy: using data that tracks single individuals over time. This strategy was first proposed by Glaeser and Mar\'e in 2001 \cite{glaeser2001cities} (see also \cite{CombesDurantonGobillon2008}). 

With data at the level of individuals, the basic idea is to compare the productivity of individual workers before and after moving to a large city. This strategy needs several additional refinements to control for other simultaneous effects that can interfere with the causal effect of city size on productivity, but makes the analysis of agglomeration economies more transparent and interesting than when one is analyzing aggregate data at the level of cities. The difficulty becomes finding such data. Data at the level of individuals is either administrative data (like insurance, tax, or social security data, which are often private and hard to access) that tend to suffer from issues of statistical representation (because these are samples from selected groups of people that are not representative of the overall population), or census data, which are statistically representative, but offer only a cross-sectional view of the population and are not longitudinal.

%%%%%%%%%%%%%%%%%%%%%%%%%%%%%%%%%%
\subsection{Agglomeration economies in developed and developing countries}
\label{sec:chAGL:empirics_countries}

Bearing in mind the methodological caveats discussed before, we review here what researchers have found about the magnitude of agglomeration economies around the world. Overall, the consensus is that the typical magnitude of agglomeration economies is approximately $4\%$ (see \cite{MeloGrahamNoland2009meta,AHLFELDT201993,grover2021agglomeration}). There is, however, great variability around the world, in part due to the methodological choices that different studies have taken. Unfortunately, most of the research in urban economics has been disproportionately focused on the developed countries. This fact can in part be explained by the wider availability of metro-level data in OECD countries.

Hence, the value of $4\%$ is mostly representative of high-income countries. In the US, for example, accepted estimates range between $3.6\%$ and $4.3\%$, while in France is around $3\%$. The few studies that have been conducted in developing countries tend to find higher elasticities. Thus, researchers have estimated elasticities of $19\%$ in China, $17\%$ in Africa, $6\%-16\%$ in Latin American countries (see \cite{MeloGrahamNoland2009meta,AHLFELDT201993,grover2021agglomeration} for recent reviews and meta-analyses of the literature). This is important, because it means that developing countries derive huge benefits from agglomeration. It urges decision makers to understand the underlying causes for the strong need in those countries for density, and the consequent pitfalls.

In summary, while researchers agree that empirical evidence supports the idea that agglomeration economies in dense urban areas boost productivity, there is a significant knowledge gap due to a lack of understanding of agglomeration economies in developing countries. Meta-analyses and literature reviews \cite{MeloGrahamNoland2009meta,AHLFELDT201993,grover2021agglomeration,duranton2020economics} rely heavily on estimates from wealthy countries, and no systematic attempt has been made to quantify the disparities in estimations between advanced and poor nations. Filling this knowledge gap is critical since the majority of people on Earth is projected to live and work in developing-country cities, which are rising quickly in population. 

%%%%%%%%%%%%%%%%%%%%%%%%%%%%%%%%%%%%%%%%%%%%%%%%%%%%%%
\section{Models and explanations from economics and complexity}
\label{sec:chAGL:smodel}

\subsection{Economics: the idea of spatial equilibrium}
\label{sec:chAGL:puzzle}
\emph{Spatial equilibrium} is the organizing idea in urban economics for thinking about cities. Here, we will focus on the inter-urban implications of spatial equilibrium (i.e., what happens across cities) rather then the intra-urban implications (i.e., what happens within cities). The literature of the former was developed by Rosen (1979) and Roback (1982), which is why economists sometimes talk about the Rosen-Roback model of urban spatial equilibrium \cite{glaeser2008cities}. 

Imagine there exists a place where the amenities are so nice that everyone wants to live there (e.g., for many of us, that would be a city with warm weather throughout the year, beautiful sunsets, fertile soil, next to both beautiful mountains and a warm beach, where people can swim in the sea and have plenty of fish they can eat with little effort). Through the exchange of information about the available options for establishing a life, people would start moving to that place. From being close to each other, the right matches between people form to create happy marriages and successful enterprises. The place would start producing amazing innovations, many of which could start being exported to the rest of the world. Wealth would accumulate, the best firms and the brightest people would want to establish themselves there, wages would increase, and all this economic success would attract more people and firms to go there. It is not hard to imagine that there is at least one place in each country on Earth that is not too far from this description. Why are we not all living in that place? Why do we live in many cities as opposed to a single (gigantic) city? What are the underlying trade-offs? Something has to give. But what? In economics, the concept of a \emph{spatial equilibrium} formalizes and captures the essential elements at play here.

Mainstream economics is centered around the idea of equilibrium - perhaps to a degree that actually hurts economic inquiry. In contrast, complexity scientists tend to think of cities as entities in disequilibrium \cite{johnson_cities_2017}. However, thinking about systems in equilibrium does help with developing some useful insights, even amid the recognition that agents do not act as perfect rational optimizers (and are, more likely, `boundedly rational' \cite{simon1955behavioral}). In urban economics, the main organizing idea is that individuals can move freely within a nation, akin to the `no friction' assumption in physics, which is an idealization that simplifies reality. It is not true in reality, given that myriad factors tether individuals to their birth cities (e.g., family, private properties they need to take care of, health, or simple lack of resources to travel), but it is a good approximation supported by the fact that migration has been one of the universals of human nature \cite{ReichDavid2018Wwaa}. Indeed, despite its costs, moving across space in search of better opportunities is one of the most effective ways to improve one's economic well-being \cite{richerson2008migration}. 

Assuming that individuals are free to move in space (between cities) implies that people are ultimately \emph{indifferent} between locations. This may seem counterintuitive, but such implication arises because we assume individuals have taken into account all benefits and costs, and have all made their final decision (i.e., if they moved and didn't like it, they are free to move again and find another location, until they are happy). Thus, the concept of a \emph{spatial equilibrium}. Let us develop some additional intuitions from this key idea.

From the point of view of individuals, three things matter (again, as a simplification): wages, rent, and amenities. When spatial equilibrium applies, an individual cannot have a good wage, a low rent, and great amenities, all three simultaneously. For example, in the beautiful city we pictured above, wages and amenities were great. However, demand to live there would make housing very expensive. Moreover, the increasing density of people may in fact damage the amenities that were the original appeal of the place (e.g., due to environmental degradation, or visual, sound, and air pollution). In this simplified version of the world, things will reach an equilibrium. In this equilibrium, not all people will decide to move to the most beautiful city because some people will choose to live in places with fewer amenities if the cost of housing is cheaper, or if wages are higher. 

The assumption of spatial equilibrium therefore implies that high wages in a place are, actually, an indication of poor amenities, high rent prices, or both. High wages and high rents do not equilibrate across cities (that is, wages and rents can continue being different between cities) because high wages in a given city are, in turn, an indication that the firms 
have decided to locate there for a reason. And the reason must be that, by locating themselves in that city, firms are being able to be more productive (and therefore more profitable).\footnote{Recall, in standard economic theory, assuming perfect competitive markets, the wage a firm pays to a worker is determined by the \emph{additional} value he or she brings to the firm. In economic language, wage is the \emph{marginal product of labor}.} So, at the end, higher wages are an indication that the area is benefiting from agglomeration.\footnote{This is formalized in a theorem first stated by Starrett in 1978 \cite{starrett1978market}, called the Spatial Impossibility Theorem, restated by Ottaviano and Thisse (see, e.g., \cite{ottaviano2004agglomeration}). The theorem implies that there are three sufficient conditions for economic activities to agglomerate: (i) heterogeneous space, (ii) positive externalities, or (iii) imperfect markets. The concept of agglomeration economies is particularly associated with (ii), but there are mechanisms in which (i) and (iii) also create agglomeration economies. We discuss the specific mechanisms in Section~\ref{sec:chAGL:causes}.}  

Through this line of reasoning we have arrived at an important insight: to reveal agglomeration economies, we should look at \emph{nominal} wages and check how they correlate with the density or the size of a city. Looking at \emph{real} wages, as opposed to nominal wages, would defeat the whole purpose. By assumption, we have said that wages are offset by the costs of living. Therefore, adjusting nominal wages to account for the costs of living would simply reveal that all real wages are equal across all locations, and we would destroy the information generated by agglomeration economies.

%%%%%%%%%%%%%%%%%%%%%%%%%%%%%%%%%%%%%%%%%%%%%%%%%%%%%%
\subsection{Economics: causes of agglomeration economies - sharing, matching and learning}
\label{sec:chAGL:causes}

Economists have been wondering for a long time about the mechanisms that would cause economic agents to benefit from crowding together in close physical proximity. The original classification for the causes of agglomeration economies is usually attributed to Alfred Marshall - a foundational figure in economics - first published in 1890 in his work \emph{Principles of Economics}.\footnote{Citations sometimes refer the 1920's 8th edition of Marshall's book.} There, Marshall proposed that agglomeration was a consequence of reducing the transportation costs for goods, people, and ideas. According to Marshall, by locating near each other, firms will reduce the shipping costs of materials (reducing, for example, the interaction costs between buyers and suppliers), the costs associated with the labor market (e.g., the search and employment of appropriate workers), and finally, the costs of finding new ideas (e.g., firms will benefit from ``knowledge spillovers'').

Duranton and Puga (2004) \cite{DurantonPuga2004micro} re-conceptualized this classification. According to them, once a firm in a particular industry is established in a city, it paves the way for other firms to get established there because the firm has assumed some initial fixed costs, has created a network of customers and suppliers, and has revealed important knowledge about how to operate successfully in the city. Thus, once the first firm arrives into a city, other firms will benefit, and this will show up as increases in productivity. In Duranton and Puga's classification, close physical proximity would entail benefits to the firms because of better 
\begin{itemize}
    \item \emph{sharing} of local infrastructure, indivisible facilities, suppliers and customers, leading to savings on fixed costs, as well as firms sharing risk, 
    \item \emph{matching} between employers and employees, or sellers and buyers, and
    \item \emph{learning} opportunities in terms of generation, spread and accumulation of knowledge.
\end{itemize} 

Since Marshall's time, the causes for the gains from agglomeration have been thoroughly researched. However, the relative impact of each of these mechanisms has been hard to quantify. This is due, in part, to the so-called ``Marshallian equivalence'' \cite{DurantonPuga2004micro}, which states that all agglomeration mechanisms provide the same prediction: firms involved in similar activity will tend to cluster since they produce benefits for one another. The addition of gains due to the three causes makes determining which mechanism has the most weight very challenging. Identifying how each of these mechanisms impact agglomeration, however, is important not only from a policy perspective, but also to anticipate how technological advances and socioeconomic change will affect the growth of cities.

The current understanding is that these mechanisms affect different industries in distinct ways, and that these effects have been changing over time. Diodato et al. (2018) \cite{diodato2018industries} carried out an extensive study of these observations. They show that value chain connections used to be dominant for most of the twentieth century for most industries. Currently, however, there has been a shift in the economy into services and skill-sharing has become the most important motivator driving firms in that sector, while value chain connections remains the main cause for co-location patterns in manufacturing.  

Are Duranton and Puga's views the final word on the origins of agglomeration economies? There are reasons to suppose that city size has more to it than their three mechanisms. For example, there are additional urban phenomena that do not represent economic advantages but are still amplified in larger cities (like diseases and crime) in the same manner that economic gains are. Yet, the mechanisms of sharing, matching, and learning are unlikely to also explain those other non-economic phenomena. From a complexity point of view, however, a simple set of mechanisms should be able to explain all these phenomena, economic and non-economic. One element that could explain why larger cities potentiate all human activities is the fact that they are more diverse (Sec.~\ref{sec:chAGL:urbanscaling} will discuss this idea).

%%%%%%%%%%%%%%%%%%%%%%%%%%%%%%%%%%%%%%%%%%%%%%%%%%%%%%
\subsection{Complex systems: a criticism of the economics approach - artificial increasing returns to scale}
\label{sec:chAGL:lognormal}
There is an empirical problem with measuring agglomeration economies that merits a section on its own here, separate from the empirical challenges that we have summarized so far. The issue we want to discuss here is one in which the statistical association that emerges between size and productivity does not arise from difficulties in disentangling economic causes and effects (i.e., cannot be represented in Fig.~\ref{fig:causaldiagram}) or from one the three economic mechanisms explained in the previous section. It is an effect from over relying on the concept of ``average'' that one often does when analyzing complex systems.

The diagram in Fig.~\ref{fig:causaldiagram} and the three mechanisms for agglomeration economies (sharing, matching and learning) both make an assumption that, at first sight, would be unobjectionable. That in order to quantify the sources of correlations between city size and productivity using data, one can safely compute \emph{averages}. For example, in Fig.~\ref{fig:wagesVSn} we looked at the \emph{average} wage across cities. Or, when researchers take care of the issues of confounding and self-sorting, they aim to measure how size will influence the productivity of individuals \emph{on average}. In all these efforts, researchers assume the Law of Large Numbers (LLN) applies: that taking the average over a finite sample of observations approximates the true underlying expected value. This assumption is fundamental to doing science in general. However, quantities in complex systems sometimes have the nasty property that their distributions either lack an expected mean, or their mean is very hard to estimate. Science in these cases must advance with caution, being mindful of this issue (see, e.g., \cite{schroeder1991fractals,Sornette2006,NewmanPowerLaws2005}). How does this problem affect the analysis of cities?

In simple words, the challenge is that cities are fundamentally unequal, heterogeneous, places. Income, productivity, wages, and other economic variables, tend to be unevenly distributed across firms and individuals in cities. The distributions that describe these quantities are often power-laws, log-normals, or more generally, heavy-tailed distributions, which limit the validity of the LLN \cite{EmbrechtsKluppelbergMikosch1997}. If the validity of the LLN is in question, one cannot take averages.

\begin{figure}[t]
%\sidecaption[t] 
\centering
\includegraphics[scale=.75]{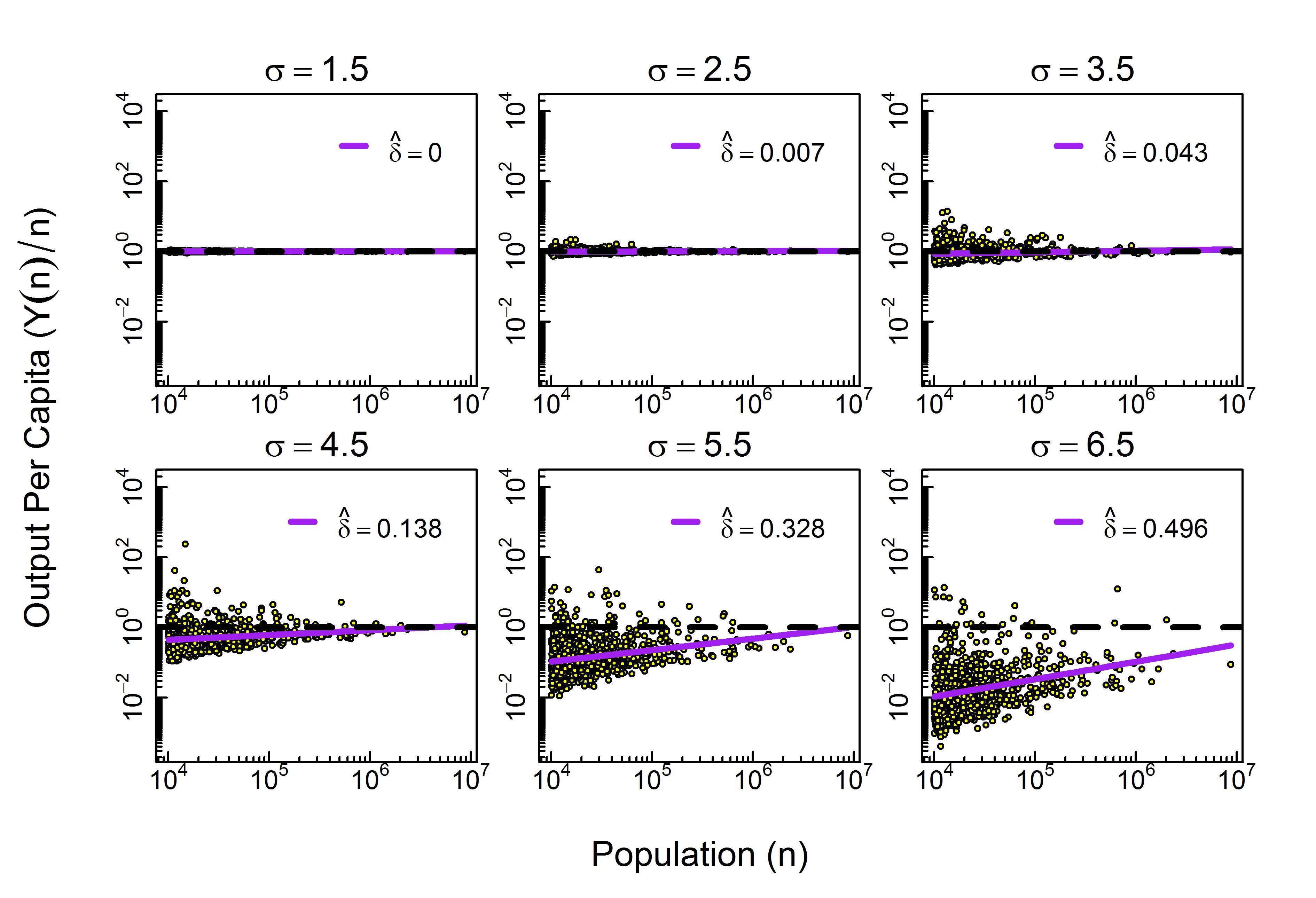}
\caption{Correlations between city size and measures of output per capita like wages emerge without agglomeration economies or self-sorting for pure statistical reasons due to wages being lognormally distributed. In these simulations, each dot represents a hypothetical city, and within each city individuals are assigned a wage sampled independent and identically from a lognormal distribution with same mean and standard deviation across all cities. Each panel plots the relationship that emerges for different standard deviations of log-wages $\sigma$. Source: figure from \cite{gomez2021artificial}.}
\label{fig:artificialIRS}       % Give a unique label
\end{figure}

To take an average of a set of observed numbers, one needs to guarantee, first, that their distribution has a finite variance, and second, that the sample size is large enough. Gomez-Lievano et al. (2021) \cite{gomez2021artificial} calculated that if urban quantities are log-normally distributed (as wages in cities tend to be), then averages can be taken only if $\ln(N)\geq \sigma^2/2$, where $N$ is the city size and $\sigma^2$ is the variance of log-wages. This relation means that the larger the variance of log-wages, the larger the size must be in order to guarantee that the sample average $\frac{1}{N}\sum_j y_j$ of wages across individuals $j=1,\ldots, N$ makes sense. An interesting effect occurs to the sample average of wages when $\ln(N)< \sigma^2/2$, that is, when LLN does not apply: the sample mean (e.g., average wage) and the sample size $N$ become positively correlated. If such situation exists in real data, higher average wages will be observed in larger cities, \emph{even if the true underlying expected mean is constant across cities}. In other words, under certain situations in which the LLN does not apply, an artificial effect emerges that can be mistakenly interpreted as evidence for agglomeration economies. However, pure independent and identically distributed random variables can generate this effect, as illustrated in Figure~\ref{fig:artificialIRS} (we encourage the reader to simulate this phenomenon and replicate this simple figure). 

This effect stands as a cautionary tale about how we think about cities. Many mathematical models assume representative agents and apply averages to data to make statements about such agents. However, the heavy-tailed distributions in cities imply that such representative agents may not exist, and that computing empirical averages may be mistaken. Conceptually, our models should instead start from the assumption that heterogeneity is a fundamental characteristic of cities. Methodologically, this effect implies that our null hypothesis should be that observing a positive association between size and productivity ought to be expected even if there are no \emph{real} causes correlating size and productivity.

%%%%%%%%%%%%%%%%%%%%%%%%%%%%%%%%%%%%%%%%%%%%%%%%%%%%%%
\subsection{Complex systems: urban scaling theory and its relationship to agglomeration economies}
\label{sec:chAGL:urbanscaling}

As has been discussed elsewhere in this book, cities have been studied using the lens of Scaling Theory. The idea behind scaling is that associations of the form $Y=a X^b$ between quantities of interest and the scale of the system may indicate a property of the phenomenon of fundamental importance: scale-invariance. That is, the rule that explains the phenomenon acts the same across scales. The success of using scaling in physics and biology to derive theories for how systems work (see, e.g., \cite{west1997general, schroeder1991fractals, barenblatt2003scaling}) encouraged the study of cities with the goal of finding a unified theory for how cities work \cite{BettencourtPLOS2010}.

Using scaling analysis, complex system scientists have observed that the relationship between many urban quantities $Y$ and size $N$ is characterized by a power-law function $(Y/Y_0)=(N/N_0)^\beta$, suggesting a universal mechanism at work (for a review of the field see Chapter 11, as well as \cite{bettencourt2021introduction}). Indeed, urban metrics following this law span several different activities and have even been observed even in archaeological data (see, e.g., \cite{ortman2014pre, lobo2020settlement}). Of particular interest for our current chapter are phenomena for which exponents $\beta$ have been observed to be larger than 1. When expressing $Y$ in per capita terms, $y=Y/N$, these \emph{super-linear} power laws indicate that larger cities have higher $y$, because $y\propto N^\delta$, with $\delta=\beta-1>0$.

What is the connection between the scaling relationships that have been identified by complexity scientists, and the relationship between productivity and population size that characterizes the phenomenon behind agglomeration economies, which is also described by a power-law (see Eq.~\ref{eq:percapitascaling})? The fact that there are other phenomena, different from income and wages, that occur with higher intensity in larger cities raises the question of whether the economic arguments described in Sections~\ref{sec:chAGL:puzzle} and \ref{sec:chAGL:causes} (ideas of spatial equilibrium, sharing, matching, or learning) are actually correct. Criminal activities, such as homicides, or the prevalence of diseases, such as cases of syphilis, also increase on a per capita basis with population size, following power-laws (see, e.g.,  \cite{GomezLievanoYounBettencourt2012,Patterson2015STDs}). Would economic arguments make sense in order to explain why homicides, or cases of syphilis, scale with size? At a first look, it does not seem likely.

Some would argue that the theory on agglomeration economies emerging from economics should be abandoned given that a more general theory may exist. Surely, a theory that can explain a large number of urban properties has a significant advantage. Alternatively, however, future work could examine the possibility that specific economic theories, such as the ones in the space of agglomeration economies, are a special case of urban scaling theory.

%%%%%%%%%%%

The promise of a unified theory of cities derived from scaling laws has not been fulfilled yet. Nevertheless, a first parsimonious explanation for the super-linear scaling of economic variables from a complex systems point of view was proposed by Bettencourt (2013) and used the idea that larger populations are inherently more interconnected \cite{Bettencourt2013}.

The argument is as follows. Assume that the output of a given individual in the city is proportional to the density of interactions that such person has with other people in the city, $y\propto \rho_\text{social}$. The density of interactions, in turn, is defined to be proportional to $N/A_\text{social}$, where population count is divided not by the surface area of the city, but by the ``social area'' in which interactions between people occur (e.g., the infrastructural network),  $A_\text{social}$. Now, let people be evenly distributed throughout the city's surface are, $A_\text{surface}$, and let the average distance between individuals be $d$. Given this, the surface area is proportional to the population size multiplied by the area between them, $A_\text{surface}\propto Nd^2$. The social area where interactions occur represents only a small fraction of the city's surface area. The model assumes that the social area is actually defined by the \emph{linear} space between people, which yields $A_\text{surface}\propto Nd$. Thus, if $\rho=N/A_\text{surface}$ is the population density, Bettencourt's assumption implies that the population density is proportional to the density of interactions squared, $\rho\propto \rho_\text{social}^2$. Going back to the output per capita, we get that 
\begin{equation}
	y\propto \rho_\text{social}\propto \rho^{1/2}
	\label{eq:popdensity}
\end{equation}
According to Equation~(\ref{eq:popdensity}), output per capita scales with the square root of city population density. In this model, this relationship holds regardless of the type of output, the physical makeup of the city, or the budget constraints of individuals. To derive a relation between $y$ and $N$ that does not depend on the area $A_\text{surface}$, however, one does need to make further assumptions. Thus, Bettencourt (2013) assumes that the minimum output per capita allowed in the city must be enough to pay for the cost of traversing the whole city. The minimum output is assumed to be proportional to the population density, $y_\text{min}\propto \rho$ (not $\rho_\text{social}$), while the cost of traversing the city is assumed to be proportional to the linear extent of the city, $c\propto L\propto A_\text{surface}^{1/2}$. The condition $y_\text{min}=c$ implies that $(N/A_\text{surface})\propto A_\text{surface}^{1/2}$, such that one can write $A_\text{surface}\propto N^{2/3}$. Putting everything together yields
\begin{equation}
	y\propto N^{1/6}.
	\label{eq:popbettencourt}
\end{equation}
In words, the model predicts an elasticity between productivity and size of $\delta=1/6\approx 16.7\%$, which is far from the estimates for developed countries we mentioned in Section~\ref{sec:chAGL:empirics_countries}, but not too far from the estimates for developing countries.

Note that Bettencourt's model is agnostic about how interactions are converted into output. The model does not indicate whether interactions represent instances of \emph{sharing} resources, or \emph{matches} between employers and employees (or suppliers and buyers), or cases of \emph{learning} due to the interchange of information between individuals when they meet. In that sense, it is more general. Note, however, that the model has similar limitations in explaining phenomena like disease or homicides. It assumes social interactions are necessary and sufficient conditions to generate the different types of output (cases of crime and disease), and it either predicts a universal relationship with population density with an exponent of $0.5$ that should hold across all activities (Eq.~(\ref{eq:popdensity})), or a relationship with city size arising from economic assumptions that involve a `minimum output per capita' equating a `cost of transportation' (Eq.~(\ref{eq:popbettencourt})). 

Some improvements on Bettencourt’s model have been proposed in which interactions are necessary but not sufficient (e.g., \cite{gomez2016explaining,gomez2021estimating}). These models also account for the cognitive constraints that limit the number of social links that humans can maintain \cite{Milgram1970, dunbar1998social, gonccalves2011modeling}, and consider individuals to be \emph{combiners} of information, rather than conduits (for a review of the evidence in this area see \cite{fogarty2015cultural}). 

We highlight the model by Gomez-Lievano et al. (2017) \cite{gomez2016explaining}, for example, in which people derive value not from the quantity of social links, but from the diversity that comes from them. In such model, individuals in larger cities are exposed to more diverse cultural factors, other people's know-how, and different pieces of information, which can be combined to generate value \cite{weitzman1998recombinant,muthukrishna2016innovation}. In this model, therefore, larger sizes lead to more diversity, and more diversity leads to higher output\footnote{It is worthwhile pointing out that this finding aligns with the concept of urbanization economies, also known as the urbanization economies as opposed to the concept of localization economies. The former involves the spatial concentration of firms due to across-industry advantages (the richness of economic activity in cities). The latter involves the clustering of similar firms due to within-industry advantages.}. Mathematically, a power-law scaling emerges as the combined outcome of exponentiating a logarithm, as $Y \propto e^{diversity} \propto e^{\beta \ln(N)+c} \propto N^\beta$. This model is well supported by theory and data (e.g., see \cite{henrich2004demography, kline2010population,YounEtAl2016Universality,zumel2020innovation}). %An analysis of the cultural creation of Rap music, which used a large data set of artists, their location, and their body of work, provided empirical support for this combinatorial hypothesis \cite{zumel2020innovation}. 
One of the predictions of this model is that socio-economic outputs that are more complex (i.e., those that require a higher diversity of inputs or ingredients to be produced) will display larger scaling exponents, will be less prevalent in general, and will be more variable from city to city. These predictions were recently verified by Balland et al. (2020) \cite{balland2020complex}.

Ribeiro and Rybski conducted a survey of the main mathematical models of urban scaling \cite{ribeiro2021mathematical}. In general, urban scaling researchers tend to depict cities as more messy, more governed by stochasticity, intricate interactions, and serendipitous complementarities among heterogeneous actors, than mainstream urban economists, who like to emphasize cities' orderly structure, the economic forces of supply and demand, and the optimal decisions from rational agents. The actual world is likely to be a blend of these different perspectives, and a discussion between academics in urban economics and complex systems science should generate useful insights.

\subsection{Complex systems: urban scaling theory criticisms}
\label{sec:chAGL:urbanscalingcritique}

Since the formulation of the urban scaling hypothesis \cite{strumsky2005metropolitan,paulus2006evolutionary,BettencourtLoboStrumsky2007}, several researchers from within the complexity field have offered criticisms of the framework and the approach. The criticisms are mostly directed at the statistical methods (see, e.g., \cite{shalizi2011scaling,leitao2016scaling,altmann2020spatial}), but some are aimed at the concept itself (e.g., \cite{arcaute2015constructing,cottineau2017diverse,sarkar2020evidence}). Literature showing counterexamples or raising contradictions in statistical analyses can be found Chapter 11. We mostly focus below on conceptual issues related to urban scaling as they relate to agglomeration economies as discussed in previous sections.

Shalizi (2011) \cite{shalizi2011scaling}, for example, argued that the relationship between economic output and city size is not a power law. Shalizi suggested that other functions of population $N$ can fit the relationship equally well. If this is true, then the relationship of urban characteristics and population is not scale invariant. Bettencourt et al. (2013) \cite{BettencourtEtAl2013hypothesis} respond to this criticism by stating that there is room for other explanations of urban scaling: ``The fact that these other functions of $N$ can fit urban data well (but not better) than power-laws in most observed cases [...] merely means that other models may be viable explanations of the scaling behavior of urban indicators, along with scale invariant functions''. When statistics is not enough to distinguish among different functional forms, the authors argue one must rely on theory: ``[B]ecause the scale-adjusted logarithmic function does not derive from theory to the best of our knowledge or fit the data manifestly better, we believe the hypothesis of urban scaling in terms of a scale invariant function, predicted by theory, remains at this point the better explanation'' (p. 12). 

Other criticisms, such as those of Leit\~ao et al. (2016) \cite{leitao2016scaling} and Altmann (2020) \cite{altmann2020spatial}, have indicated that a misspecification of the distribution of the dependent variable in urban scaling analysis, or the independence assumption between observations conventionally made in linear regressions, lead to wrong conclusions about the functional form relating outcomes $Y$ with city sizes $N$. These are valid concerns, and researchers will need to rely on theory to decide which statistical methods and functions to use to fit the data, as Bettencourt et al. (2013) \cite{BettencourtEtAl2013hypothesis} argue. 

To these methodological criticisms, we can add the empirical challenges we mentioned in Section~\ref{sec:chAGL:challenges}, in which the statistical approaches used by urban scaling researchers to estimate the power-law exponents do not address the issues of confounders and reverse causality. Complexity scientists might argue that they are not concerned about reverse causality, since the question in urban scaling is not necessarily why size \emph{causes} productivity, but why size and productivity maintain a very special relationship with one another across scales. The problem of confounders, however, still holds, since confounding adds spurious correlations between variables (see \cite{Patterson2015STDs} for an example of urban scaling analysis that does take into account confounders). 

But what if the very premise of urban scaling analysis, that aggregate urban quantities are a function of total size, is mistaken? Such conceptual questions started with Arcaute et al. (2015) \cite{arcaute2015constructing}, who noted that cities are not objects with clear physical boundaries, and therefore there is no definitive way of quantifying their size, or many other urban quantities. Thus, the premise of urban scaling may be misplaced. These authors showed how different ways of delineating the boundaries of cities return power-laws with different exponents, putting in doubt the very concept of agglomeration economies. Other studies have followed (e.g., \cite{cottineau2017diverse,sarkar2020evidence}) arguing that researchers should be more cautious about computing aggregate measures like total population size, or total output. Instead, they should include in their models the specific intra-city composition of economic and social
activity and physical infrastructure in order to understand urban economic performance.  

Note that the criticisms raised above have been addressed by mainstream urban economists by having a strict definition of cities as ``labor markets'' (i.e., places were people live and work), and by focusing their empirical work to studying the effects of agglomeration at the micro level of individual firms and workers. Given this focus, both their mathematical models and their statistical methods can be fairly detailed about mechanisms, causes and effects. %It seems that both complex systems scientists and economists should take more seriously the study of cities as complex systems and analyze them instead as processes with an extension in time and space and complex internal structures.

%%%%%%%%%%%%%%%%%%%%%%%%%%%%%%%%%%%%%%%%%%%%%%%%%%%%%%
\section{Complexity and economics: convergence for a better understanding of cities?}
\label{sec:chAGL:complexity}

Since at least the 1980s, several efforts of bringing a diverse set of economists, physicists and other scholars in fields related to complexity science have produced important research \cite{andersonarrowpines1988economy}. The field of complexity economics emerged from that work: assumptions of perfect rationality, well-defined problems and optimal behavior were relaxed. Heterogeneous agents, each with imperfect information and capable of adaptation to their environments were introduced. The approach brought about models of economies that are ``organic, always creating itself, alive and full of messy vitality.'' \cite{arthur2021foundations}. Unfortunately, its impact in the mainstream of the economics profession is debatable. 

The original challenge for the new field involved established research ``cultures'' and methodological or modeling choices or traditions. Mainstream economists have been modeling social phenomena utilizing representative agents (e.g. households, firms and developers) who solve optimization problems with the end outcome of the system arriving at an equilibrium. Complexity science-insprired economics models social phenomena utilizing heterogeneous agents, with localized knowledge, following rules and adapting to strategies of neighbors. While both approaches lead to emergent aggregate phenomena, the complexity economics approach is broader in scope and does not rely on limiting assumptions made in mainstream economics (representative agents, rationality, complete markets, etc). Below we summarize and contrasts the major features typically found in both mainstream urban economics and the complexity-oriented urban scaling theory approaches (Table \ref{tab:contrast}).

\begin{table}[h]
\caption {Summary of features of urban economics and urban scaling theory approaches} \label{tab:contrast} 
\begin{tabularx}{\textwidth}{l|X|X}
\hline
\textit{Feature} & \textit{Urban Economics} & \textit{Urban Scaling} \\
\hline
Agents & Representative (but N different types) & Diverse \\ 
Behavior & Optimizing (depending on agent type - utility, profit, rents) & No goals other than interacting locally \\
Rationality & Perfect & Not defined\\
Interactions & Homogeneous (impersonal markets, no network formation) & Channeled by networks (social and physical) of differing strength and types\\
Interaction extent & Global & Localized\\
Theme & Optimal resource allocation and spatial equilibrium & Formation of structures, emergence of socioeconomic aggregates\\
Structural change & Equilibrium shifts & Endogenous re-organization\\ 
\hline
\end{tabularx}
\text{Source: \cite{arthur2021foundations} (table adapted by the authors)\\}
\end{table}

%\end{center}

Thisse (2014) \cite{thisse2014new} offers one of the few attempts made from mainstream economists, exploring the possibility of a synthesis between the field of urban economics and urban geography (which embodies several themes of complexity). Reviewing the work by Batty (2013) \cite{batty2013new}, he reaches a few  conclusions that are revealing. Thisse focuses on the fundamental aspect of the methodological choices an economist makes and the associated belief system: `` macroscopic phenomena can best be understood by examining how they arise from the motivated actions and interactions of individual agents in a world where various types of uncertainty prevail.'' This is not a point that a complexity scientist would contest. But he takes issue with an urban complexity view ``formed by purposeless agents whose aggregate behavior is governed by stochastic laws.'' (Mainstream) economists model cities ``as emerging from the decisions of many rational agents—firms, households, developers, and local governments—whose actions are often coordinated by markets.'' He correctly suggests that this difference in approaches is not ``an insurmountable obstacle to interactions between urban geographers and urban economists''.

While interactions can occur without a significant efforts for unifying our understanding of the functioning of cities, we argue that what is needed is parallel shifts in urban modeler methodological positions. For example, on the one hand, economists need to rely less on representative agents and rational expectations and allow for agent heterogeneity. The strong emphasis on microfoundations leading to macro-outcomes has required unrealistic assumptions to the detriment of whole fields of study in economics \cite{stiglitz2018modern}. Furthermore, network structure of interactions between people is important. Consumer and producers do not only interact in anonymous market through the price system, but develop relationships of differing strength in networks. Also, moving away from analytical models and adopting agent-based simulation modeling approaches in the context of economics could help bridge some of the gaps that exist. Some progress towards this goals already exists (\cite{behrens2015agglomeration}) but additional efforts will assist in the convergence of these knowledge domains.

On the other hand, urban scaling researchers (complexity scientists, physicists, statisticians, urban geographers, etc.) should attempt to further elaborate the existence of markets and distinct mechanisms of production and exchange in their models. The emergence of markets or other economic institutions from interactions within population (who adopt roles of producers and consumers) would be even more valuable. We are also faced with a paradox in urban scaling theory. In some ways, the theory appears to not allow for enough depth in idiosyncratic dimensions. Some people are more talented than others and could be attracted to bigger cities based on their individual characteristics. Some cities become diversified and some specialized in terms of their economic base. The migration of population from rural to urban areas is also not part of urban scaling theory. Business cycles - periods of economic booms and busts - are not explained. This richness of the urban experience is currently somewhat lost in a general theory of urbanization. Several attempts to close this gap are underway (e.g., \cite{gomez2016explaining,pumain2018evolutionary,bettencourt2019towards,gomez2021estimating}).

Finally, it's worthwhile pointing out that differences between the two fields are also due to the way that research questions are asked. In urban scaling theory, the main question asked is whether the scaling of urban phenomena (an aggregate outcome) can be attributed to localized agent interactions (micro behavior). In economics, the questions are much more focused and directed: what type of economies do we observe? what are their causes within the operation of a market system? If scaling theory is indeed a `General Theory of Urbanization', we should have observed similar scaling phenomena across distinct economic structures (for example, were systems of cities in the USSR experiencing similar scaling phenomena?) Some evidence exists on scaling of settlements in ancient societies - which were admittedly much simpler in terms of the functioning of markets, exchange and production. We argue that more research is needed when considering the potentially differential effects of modern economic structure.

%%%%%%%%%%%%%%%%%%%%%%%%%%%%%%%%%%%%%%%%%%%%%%%%%%%%%%
\section{Assessing agglomeration effects beyond economics and complexity}\label{sec:chAGL:beyond}

We have summarized the topic of agglomeration economies from the economics and the complex systems perspectives. However, economists and complexity scientists are not the only ones who have studied the effects of agglomeration, and these other perspectives merit a brief mention.

We pointed out in Section~\ref{sec:chAGL:causes} that the economist Alfred Marshall had speculated about the causes and the effects of physical proximity back in the 1890's. He wrote about this question after noticing that industries specialize geographically. As we explained, he contended that spatial proximity promotes the transmission of knowledge, reduces transport costs of inputs and outputs, and allows firms to benefit from a more efficient labor market. At around the same time, the French sociologist \'{E}mile Durkheim also remarked the effects of agglomeration from a sociological point of view. These and many other of Durkheim's ideas were published as part of his 1893 doctoral dissertation, \emph{The Division of Labour in Society} \cite{durkheim1893division}. Durkheim is seldom cited by economists and complexity scientists, but his influence might be as important, or more, than Marshall's. It is interesting to observe that the central issue of population scale and density have been addressed by foundational figures in both economics and sociology. 

While Marshall emphasized the \emph{immediate} effects of agglomeration on \emph{firms}, Durkheim underlined the \emph{collective} effects on \emph{society}. In Durkheim's view, population density and growth were the crucial drivers of social and economic development. Durkheim's theory is interesting in that increasing the number of people in a location would increase the number of social interactions, but he did not interpret the latter to lead directly to prosperity. Instead, in Durkheim's account, the increase in the number of interactions would lead to more competition, division of labor, specialization, diversity, and ultimately, conflict. These, in turn, would force society to develop norms and behaviors to generate, sustain, and improve social integration. This vision was encapsulated in Durkheim's concept of ``dynamic density''.

Durkheim's ideas were further developed by the American sociologist Louis Wirth (see \cite{wirth1938urbanism}), who wrote explicitly about cities, with clarity and insight. These ideas, however, were not formalized into mathematical models until the 1970s. Once sociologists started to connect models with data, a heated debate was sparked about the effects of urban agglomeration (see, e.g., \cite{Milgram1970,kasarda1974structural,kasarda1979ratio,Mayhew1976,Fischer1977comment,seiyama1986review,rotolo2006population}). On the one hand, some researchers contended that Durkheim and Wirth's theories were unnecessarily complicated. The fact that social interactions scale as the square of population size, $I\propto N^2$, was enough to explain the positive relationships between socio-economic rates and the size of cities. Mayhew and Levinger (1976) \cite{Mayhew1976}, for example, used this argument to explain the higher rates of crime in large cities. On the other, such mathematical arguments were viewed as too simplistic, missing ``substantive understanding'' of how social systems work. It is instructive to read, for example, the criticism raised by the sociologist Claude S. Fischer (1977) \cite{Fischer1977comment}. For Fischer, in brief, the fact that \emph{potential} interactions grow super-linearly with size does not imply that \emph{actual} interactions do (``why do they [the proponents of the interactions-based theory] not predict increasing marriage or birth rates with city size?'', p.~453). Fischer condemns such models, because they ``assume that people in cities are like numbered marbles in a large urn, constantly being shaken up and randomly `contacting' one another. [These models are] not even a heuristically useful approximation of the truth. Casual reading of urban ethnographies---even a casual look out the window---will show that urbanites' lives cannot be described this way. Most city dwellers lead sensible, circumscribed lives, rarely go downtown, hardly know areas of the city they neither live nor work in, and see (in any sociologically meaningful way) only a tiny fraction of the city's population.'' 

Finally, a similar debate about the effects of population size on social systems has taken place, almost in parallel and with no connection to economics, complexity, or sociology, in the fields of cultural and evolutionary anthropology, where the issue is referred to as the ``population size hypothesis'' (see, e.g., \cite{carneiro1967relationship,boserup1981population,lee1988induced,shennandemography2001,henrich2004demography,kline2010population,collard2013population}). In anthropology, one of the central concerns is to explain how cultures have evolved, allowing humans to occupy almost all ecological habitats on Earth (without subspeciation, which is the typical mechanism for all other organisms to adapt to new environments) \cite{henrich2015secret}. Given the limits to what they can measure from archaeological and ethnographic data, anthropologists have studied the ``cultural complexity'' of societies (rather than productivity), defined by them as the number of cultural traits (literally, the count of distinct artefacts, social roles, etc). 

When data started to accumulate enough to be amenable to statistical analysis, a clear relationship between population size and cultural complexity across several societies was found in 1960s \cite{carneiro1967relationship}.\footnote{The anthropologist Robert Carneiro was a strong supporter of adapting insights from the physical and natural sciences into anthropology. Incidentally, he introduced an idea from psychology called ``scale analysis'' \cite{carneiro1962scale}, which should not be confused with ``scaling analysis'' as used in the urban scaling literature. His methodologies pre-dated the development of the field of economic complexity by almost 50 years (see, e.g., \cite{HidalgoHausmann2009}).} That is, societies that had larger population sizes were observed to have more technologies, deeper divisions of labor, and more political, religious, and organizational units (for recent analysis of this association see \cite{kline2010population,collard2013population}). Since then, the question has been whether population size leads to higher rates of cultural innovation, vice versa, or whether there are other variables causing both in tandem (i.e., the same questions we ask in Fig.~\ref{fig:causaldiagram}).

One of the current theories to explain why larger population sizes give rise to higher rates of cultural complexity is Joseph Henrich's ``collective brain hypothesis'' \cite{muthukrishna2016innovation}. The argument is very different from those of economists, complexity scientists, or sociologists that we have reviewed so far in that Henrich's explanation is rooted in evolutionary theory (the field itself is called ``cultural evolution''). The basis on which this theory is built are the recent observations that show that humans are super-imitators (for a review, see \cite{Laland2017}). This unique cognitive feature, unmatched across the animal kingdom, allows innovations to spread and pass down across generations without deterioration \cite{henrich2004demography}. As a result of this unique ability to mimic our close peers, cultures can accumulate an ever-expanding repertory of practices, conventions, tools, and know-how. The cultural brain hypothesis suggests that as culture evolves and expands, it exerts evolutionary forces that adjust sociality, imitation, and transmission variance to cope with the increased complexity of tools, practices, beliefs, and behaviors. In this view, the characteristics of individuals remain more or less constant across populations of different sizes, but not the ``collective brain'' of those societies. Instead, collective brains get larger as populations get larger. Hence, agglomeration leads to more innovation, higher productivity, and even higher IQ scores, but not because of smarter individual brains, but because of smarter collective brains. %Indeed, evidence in support of this argument shows that larger populations with more complex technologies engage in more teaching \cite{Laland2017}.

How do these multi-disciplinary perspectives enrich our understanding of agglomeration economies in cities? Complexity scientists ought to be in the position to synthesize the key ideas and propose new ones. We end this chapter with some final conclusions and questions for further thought.  

%%%%%%%%%%%%%%%%%%%%%%%%%%%%%%%%%%%%%%%%%%%%%%%%%%%%%%
\section{Conclusions and open questions}
\label{sec:chAGL:conclusions}

We argue that understanding cities is a very important priority for humanity, as we increasingly become an urban civilization. This understanding passes through the understanding of both benefits and costs of agglomerating.

Researchers exploring this fundamental aspect of the function of cities should be aware of the multiplicity of issues related to measurement of productivity and scale (population size vs. population density), as well as the definition of the boundaries of a city or a metropolitan area. Conducting sensitivity analysis across alternative measurement choices should be an aim of empirical research.

The economics of density, both in terms of the benefits through agglomeration economies, and the costs - congestion, crime, etc. - can be viewed both complimentary but also as a subset of the work occurring through complexity science. Agglomeration economies, from the economics perspective, is a phenomenon that relates to firms and the production of goods and services. The explanations offered by scaling theory could be viewed as a more abstract level of explanation of agglomeration economies; still, explanations offered by mainstream economists add a level of detail that is not addressed yet in the broader scaling context. Economic theory conceptualizes agglomeration economies as an economic production and cost specific concept while urban scaling theory discusses it as a broader concept, with application beyond the economics conceptualization. Unifying the approaches will, in the foreseeable future, be a challenge due to methodological differences between the fields - mainstream economists are averse to the idea of agents who are not driven by some degree of self interest and the lack of modeling of markets and exchange.

Perhaps what would be beneficial for a possible unification would be that the complexity approach adopts some micro-foundations that align with the economics approach. This could be amended by developing spatially- and network-explicit agent-based models of urban economies across scales- modeling consumption and production processes within and across cities as well as other dimensions of urban life. These models are expected to produce emergent power/scaling laws for various urban phenomena in systems of cities. In this way, explanations of the observed phenomena across the distinct modeling traditions could be more directly linked.

Urban scaling theory has done a substantial amount of work on the costs of scale or density - what Glaeser (2011) called the ``demons of density''. While some of the obvious challenges were identified early in the economics literature \cite{pressman1971crime}, comparatively, very little attention has been placed on the microfoundations of costs of agglomeration as compared to the benefits of agglomeration. Complementarities could be present in this space, and urban scaling theory could offer guidance. More attention should be placed in uniting social economics theories to those of urban economics, perhaps supplemented by social psychology theories on crowding.

It is worthwhile noting that as more detailed models of urban scaling come about, the connections between the complexity approach and other fields can begin to emerge. For example, modeling people as derive value not from the quantity of social links, but from the diversity that comes from them \cite{gomez2016explaining} aligns with the concept of urbanization economies, emerging in part from what is known as Jacobs knowledge spillovers (for a review of evidence for Jacobs externalities, see \cite{beaudry2009s}). Still, urban scaling researchers need to address the concept of localization economies and identify whether scaling theory has a contribution to make in the discussion of localization vs urbanization economies.

The topic of this chapter presents several opportunities for novel research and synthesis of the existing approaches. Thus, we end with some open questions and proposed topics for future research.

\vspace{1cm}
\textbf{Questions for further discussion}
\begin{itemize}
    \item Are the causes of agglomeration economies as explained by mainstream economists a subset of a complexity science based urban scaling theory? For example, can we express sharing, matching and learning as production theory concepts that are special cases of broader concepts in scaling theory? Would those be: quantity of linkages, quality of linkages and adaptation? Or other features of scaling/\-complexity theory? 
    \item Can urban scaling theory incorporate distinct types of agents interacting in networks in a dynamic setting, through evolving norms and institutions? E.g. firm formation, land development activity, evolving markets and economic institutions, etc. 
    \item What are the possible pathways through which mainstream urban economists can integrate concepts from complexity science into their theory and models?
    \item How to enrich our understanding of the advantages of agglomeration as arising from evolutionary processes that involve history and path-dependence?
\end{itemize}

%======================================================================%
% Bibliography
%======================================================================%

\bibliography{Refs.bib}

\begin{thebibliography}{10}
\providecommand{\url}[1]{{#1}}
\providecommand{\urlprefix}{URL }
\expandafter\ifx\csname urlstyle\endcsname\relax
  \providecommand{\doi}[1]{DOI~\discretionary{}{}{}#1}\else
  \providecommand{\doi}{DOI~\discretionary{}{}{}\begingroup
  \urlstyle{rm}\Url}\fi

\bibitem{AHLFELDT201993}
Ahlfeldt, G.M., Pietrostefani, E.: The economic effects of density: A
  synthesis.
\newblock Journal of Urban Economics \textbf{111}, 93--107 (2019).
\newblock \doi{https://doi.org/10.1016/j.jue.2019.04.006}

\bibitem{altmann2020spatial}
Altmann, E.G.: Spatial interactions in urban scaling laws.
\newblock Plos one \textbf{15}(12), e0243390 (2020)

\bibitem{andersonarrowpines1988economy}
Anderson, P.W., Arrow, K.J., , Pines, D.: The economy as an evolving complex
  system.
\newblock Addison-Wesley, Redwood City, CA (1988)

\bibitem{arcaute2015constructing}
Arcaute, E., Hatna, E., Ferguson, P., Youn, H., Johansson, A., Batty, M.:
  Constructing cities, deconstructing scaling laws.
\newblock Journal of the royal society interface \textbf{12}(102), 20140745
  (2015)

\bibitem{arthur2021foundations}
Artur, W.: Foundations of complexity economics.
\newblock Nature Reviews: Physics \textbf{3}, 136--145 (2021)

\bibitem{balland2020complex}
Balland, P.A., Jara-Figueroa, C., Petralia, S.G., Steijn, M., Rigby, D.L.,
  Hidalgo, C.A.: Complex economic activities concentrate in large cities.
\newblock Nature human behaviour \textbf{4}(3), 248--254 (2020)

\bibitem{barenblatt2003scaling}
Barenblatt, G.I.: Scaling, vol.~34.
\newblock Cambridge University Press (2003)

\bibitem{batty2013new}
Batty, M.: The new science of cities.
\newblock MIT press (2013)

\bibitem{johnson_cities_2017}
Batty, M.: Cities in disequilibrium.
\newblock In: J.~Johnson, A.~Nowak, P.~Ormerod, B.~Rosewell, Y.C. Zhang (eds.)
  Non-Equilibrium Social Science and Policy, pp. 81--96. Springer International
  Publishing (2017).
\newblock
  \urlprefix\url{http://link.springer.com/10.1007/978-3-319-42424-8\_6}.
\newblock Series Title: Understanding Complex Systems

\bibitem{beaudry2009s}
Beaudry, C., Schiffauerova, A.: {Who's right, Marshall or Jacobs? The
  localization versus urbanization debate}.
\newblock Research policy \textbf{38}(2), 318--337 (2009)

\bibitem{behrens2015agglomeration}
Behrens, K., Robert-Nicoud, F.: Agglomeration theory with heterogeneous agents.
\newblock Handbook of regional and urban economics \textbf{5}, 171--245 (2015)

\bibitem{bettencourt2019towards}
Bettencourt, L.M.: Towards a statistical mechanics of cities.
\newblock Comptes Rendus Physique \textbf{20}(4), 308--318 (2019)

\bibitem{bettencourt2021introduction}
Bettencourt, L.M.: Introduction to urban science: evidence and theory of cities
  as complex systems.
\newblock MIT Press (2021)

\bibitem{Bettencourt2013}
Bettencourt, L.M.A.: {The Origins of Scaling in Cities}.
\newblock {Science} \textbf{340}(6139), 1438--1441 (2013).
\newblock \doi{10.1126/science.1235823}

\bibitem{BettencourtLoboStrumsky2007}
Bettencourt, L.M.A., Lobo, J., Strumsky, D.: {Invention in the city: Increasing
  returns to patenting as a scaling function of metropolitan size}.
\newblock Research Policy \textbf{36}(1), 107--120 (2007).
\newblock \doi{10.1016/j.respol.2006.09.026}.
\newblock
  \urlprefix\url{http://www.sciencedirect.com/science/article/pii/S0048733306001661}

\bibitem{BettencourtPLOS2010}
Bettencourt, L.M.A., Lobo, J., Strumsky, D., West, G.B.: {Urban Scaling and Its
  Deviations: Revealing the Structure of Wealth, Innovation and Crime across
  Cities}.
\newblock PLoS ONE \textbf{5}(11), e13541 (2010).
\newblock \doi{10.1371/journal.pone.0013541}

\bibitem{BettencourtEtAl2013hypothesis}
Bettencourt, L.M.A., Lobo, J., Youn, H.: {The hypothesis of urban scaling:
  formalization, implications and challenges} (2013).
\newblock ArXiv:1301.5919v1 [physics.soc-ph] 24 Jan 2013

\bibitem{boserup1981population}
Boserup, E.: Population and technological change: A study of long-term trends.
\newblock University of Chicago Press, Chicago Ill, United States (1981)

\bibitem{BEApersonalincomedata}
{Bureau of Economic Analysis (BEA)}: {To access the data, go to ``PERSONAL
  INCOME AND EMPLOYMENT BY COUNTY AND METROPOLITAN AREA'', click ``Personal
  Income, Population, Per Capita Personal Income (CAINC1)'', select
  ``Metropolitan Statistical Area'' and download the data for ``All areas''.
  Last accessed: May 15, 2022.}
\newblock
  \urlprefix\url{https://apps.bea.gov/iTable/iTable.cfm?reqid=70\&step=1\&acrdn=6}

\bibitem{BLSwages2020}
{Bureau of Labor Statistics (BLS)}: {Occupational Employment and Wage
  Statistics (OEWS), May 2020, `Metropolitan and nonmetropolitan area'. Last
  accessed: May 15, 2022.}
\newblock \urlprefix\url{https://www.bls.gov/oes/special-requests/oesm20ma.zip}

\bibitem{carneiro1962scale}
Carneiro, R.L.: Scale analysis as an instrument for the study of cultural
  evolution.
\newblock Southwestern Journal of Anthropology \textbf{18}(2), 149--169 (1962)

\bibitem{carneiro1967relationship}
Carneiro, R.L.: On the relationship between size of population and complexity
  of social organization.
\newblock Journal of Anthropological Research \textbf{42}(3), 355--364 (1967)

\bibitem{collard2013population}
Collard, M., Buchanan, B., O'Brien, M.J.: Population size as an explanation for
  patterns in the paleolithic archaeological record: more caution is needed.
\newblock Current Anthropology \textbf{54}(S8), S388--S396 (2013)

\bibitem{CombesDurantonGobillon2008}
Combes, P.P., Duranton, G., Gobillon, L.: Spatial wage disparities: Sorting
  matters!
\newblock Journal of Urban Economics \textbf{63}(2), 723--742 (2008)

\bibitem{Combes2011identification}
Combes, P.P., Duranton, G., Gobillon, L.: {The identification of agglomeration
  economies}.
\newblock Journal of Economic Geography \textbf{11}(2), 253--266 (2011)

\bibitem{combes2010estimating}
Combes, P.P., Duranton, G., Gobillon, L., Roux, S.: Estimating agglomeration
  economies with history, geology, and worker effects.
\newblock In: Agglomeration economics, pp. 15--66. University of Chicago Press
  (2010)

\bibitem{CombesGobillon2015handbook}
Combes, P.P., Gobillon, L.: Chapter 5 - the empirics of agglomeration
  economies.
\newblock In: J.V.H. Gilles~Duranton, W.C. Strange (eds.) Handbook of Regional
  and Urban Economics, \emph{Handbook of Regional and Urban Economics}, vol.~5,
  pp. 247 -- 348. Elsevier (2015).
\newblock \doi{https://doi.org/10.1016/B978-0-444-59517-1.00005-2}.
\newblock
  \urlprefix\url{https://www.sciencedirect.com/science/article/pii/B9780444595171000052}

\bibitem{cottineau2017diverse}
Cottineau, C., Hatna, E., Arcaute, E., Batty, M.: Diverse cities or the
  systematic paradox of urban scaling laws.
\newblock Computers, environment and urban systems \textbf{63}, 80--94 (2017)

\bibitem{diodato2018industries}
Diodato, D., Neffke, F., O'Clery, N.: {Why do industries coagglomerate? How
  Marshallian externalities differ by industry and have evolved over time}.
\newblock {Journal of Urban Economics} \textbf{106}, 1--26 (2018)

\bibitem{dunbar1998social}
Dunbar, R.I.: The social brain hypothesis.
\newblock Evolutionary Anthropology: Issues, News, and Reviews: Issues, News,
  and Reviews \textbf{6}(5), 178--190 (1998)

\bibitem{DurantonPuga2004micro}
Duranton, G., Puga, D.: Micro-foundations of urban agglomeration economies.
\newblock Handbook of regional and urban economics \textbf{4}, 2063--2117
  (2004)

\bibitem{duranton2020economics}
Duranton, G., Puga, D.: The economics of urban density.
\newblock Journal of economic perspectives \textbf{34}(3), 3--26 (2020)

\bibitem{durkheim1893division}
Durkheim, {\'E}.: De la division du travail social.
\newblock {F\'elix Alcan}, Paris (1893)

\bibitem{EmbrechtsKluppelbergMikosch1997}
Embrechts, P., Kl{\"u}ppelberg, C., Mikosch, T.: Modelling extremal events: for
  insurance and finance, vol.~33.
\newblock Springer (1997)

\bibitem{felipe2003aggregation}
Felipe, J., Fisher, F.M.: Aggregation in production functions: what applied
  economists should know.
\newblock Metroeconomica \textbf{54}(2-3), 208--262 (2003)

\bibitem{Fischer1977comment}
Fischer, C.S.: {Comment on Mayhew and Levinger's ``Size and the Density of
  Interaction in Human Aggregates''} (1977)

\bibitem{fogarty2015cultural}
Fogarty, L., Creanza, N., Feldman, M.W.: Cultural evolutionary perspectives on
  creativity and human innovation.
\newblock Trends in ecology \& evolution \textbf{30}(12), 736--754 (2015)

\bibitem{fujita1999spatial}
Fujita, M., Krugman, P.R., Venables, A.: {The Spatial Economy: Cities, Regions,
  and International Trade}.
\newblock MIT press (1999)

\bibitem{glaeser2008cities}
Glaeser, E.L.: Cities, agglomeration, and spatial equilibrium.
\newblock OUP Oxford (2008)

\bibitem{Glaeser2011Totc}
Glaeser, E.L.: Triumph of the city : how our greatest invention makes us
  richer, smarter, greener, healthier, and happier.
\newblock Penguin Press, New York (2011)

\bibitem{glaeser2001cities}
Glaeser, E.L., Mare, D.C.: Cities and skills.
\newblock {Journal of Labor Economics} \textbf{19}(2), 316--342 (2001)

\bibitem{goldstein1984economies}
Goldstein, G.S., Gronberg, T.J.: Economies of scope and economies of
  agglomeration.
\newblock Journal of Urban Economics \textbf{16}(1), 91--104 (1984)

\bibitem{gomez2021estimating}
Gomez-Lievano, A., Patterson-Lomba, O.: Estimating the drivers of urban
  economic complexity and their connection to economic performance.
\newblock Royal Society open science \textbf{8}(9), 210670 (2021)

\bibitem{gomez2016explaining}
Gomez-Lievano, A., Patterson-Lomba, O., Hausmann, R.: Explaining the
  prevalence, scaling and variance of urban phenomena.
\newblock Nature Human Behaviour \textbf{1}, 0012 (2017)

\bibitem{gomez2021artificial}
G{\'o}mez-Li{\'e}vano, A., Vysotsky, V., Lobo, J.: Artificial increasing
  returns to scale and the problem of sampling from lognormals.
\newblock Environment and Planning B: Urban Analytics and City Science
  \textbf{48}(6), 1574--1590 (2021)

\bibitem{GomezLievanoYounBettencourt2012}
Gomez-Lievano, A., Youn, H., Bettencourt, L.M.A.: {The Statistics of Urban
  Scaling and Their Connection to Zipf's Law}.
\newblock PLoS ONE \textbf{7}(7), e40393 (2012).
\newblock \doi{10.1371/journal.pone.0040393}.
\newblock \urlprefix\url{http://dx.plos.org/10.1371/journal.pone.0040393}

\bibitem{gonccalves2011modeling}
Gon{\c{c}}alves, B., Perra, N., Vespignani, A.: Modeling users' activity on
  twitter networks: Validation of dunbar's number.
\newblock PloS one \textbf{6}(8), e22656 (2011)

\bibitem{grover2021agglomeration}
Grover, A., Lall, S.V., Timmis, J.: {Agglomeration Economies in Developing
  Countries}.
\newblock Policy research working papers, {World Bank, Washington, DC} (2021).
\newblock \urlprefix\url{http://hdl.handle.net/10986/36003}

\bibitem{henrich2004demography}
Henrich, J.: Demography and cultural evolution: how adaptive cultural processes
  can produce maladaptive losses: the tasmanian case.
\newblock American Antiquity pp. 197--214 (2004)

\bibitem{henrich2015secret}
Henrich, J.: The secret of our success: how culture is driving human evolution,
  domesticating our species, and making us smarter.
\newblock Princeton University Press (2015)

\bibitem{HidalgoHausmann2009}
Hidalgo, C.A., Hausmann, R.: The building blocks of economic complexity.
\newblock {PNAS} \textbf{106}(25), 10570--10575 (2009).
\newblock \doi{10.1007/s10887-011-9071-4}

\bibitem{kasarda1974structural}
Kasarda, J.D.: The structural implications of social system size: A three-level
  analysis.
\newblock American Sociological Review pp. 19--28 (1974)

\bibitem{kasarda1979ratio}
Kasarda, J.D., Nolan, P.D.: Ratio measurement and theoretical inference in
  social research.
\newblock Social Forces \textbf{58}(1), 212--227 (1979)

\bibitem{kline2010population}
Kline, M.A., Boyd, R.: Population size predicts technological complexity in
  oceania.
\newblock Proceedings of the Royal Society B: Biological Sciences
  \textbf{277}(1693), 2559--2564 (2010)

\bibitem{krugman1991increasingreturns}
Krugman, P.: Increasing returns and economic geography.
\newblock {Journal of Political Economy} \textbf{99}, 483--499 (1991)

\bibitem{Laland2017}
Laland, K.N.: Darwin's unfinished symphony: how culture made the human mind.
\newblock Princeton University Press, Princeton (2017)

\bibitem{lee1988induced}
Lee, R.D.: Induced population growth and induced technological progress: Their
  interaction in the accelerating stage.
\newblock Mathematical Population Studies \textbf{1}(3), 265--288 (1988)

\bibitem{leitao2016scaling}
Leitao, J.C., Miotto, J.M., Gerlach, M., Altmann, E.G.: Is this scaling
  nonlinear?
\newblock Royal Society open science \textbf{3}(7), 150649 (2016)

\bibitem{lobo2020settlement}
Lobo, J., Bettencourt, L.M., Smith, M.E., Ortman, S.: Settlement scaling
  theory: Bridging the study of ancient and contemporary urban systems.
\newblock Urban Studies \textbf{57}(4), 731--747 (2020)

\bibitem{Mayhew1976}
Mayhew, B.H., Levinger, R.L.: {Size and the Density of Interaction in Human
  Aggregates}.
\newblock American Journal of Sociology pp. 86--110 (1976)

\bibitem{MeloGrahamNoland2009meta}
Melo, P.C., Graham, D.J., Noland, R.B.: A meta-analysis of estimates of urban
  agglomeration economies.
\newblock {Regional Science and Urban Economics} \textbf{39}(3), 332--342
  (2009)

\bibitem{Milgram1970}
Milgram, S.: The experience of living in cities.
\newblock Science \textbf{167}(3924), 1461--1468 (1970).
\newblock \urlprefix\url{http://www.jstor.org/stable/1728966}

\bibitem{moomaw1981productivity}
Moomaw, R.L.: Productivity and city size: a critique of the evidence.
\newblock The Quarterly Journal of Economics \textbf{96}(4), 675--688 (1981)

\bibitem{muthukrishna2016innovation}
Muthukrishna, M., Henrich, J.: Innovation in the collective brain.
\newblock Philosophical Transactions of the Royal Society B: Biological
  Sciences \textbf{371}(1690), 20150192 (2016)

\bibitem{NewmanPowerLaws2005}
Newman, M.E.J.: {Power laws, Pareto distributions and Zipf's law}.
\newblock Cont. Phys. \textbf{46}(5), 323--351 (2005).
\newblock \doi{10.1080/00107510500052444}

\bibitem{ortman2014pre}
Ortman, S.G., Cabaniss, A.H., Sturm, J.O., Bettencourt, L.M.: The pre-history
  of urban scaling.
\newblock PloS one \textbf{9}(2), e87902 (2014)

\bibitem{sullivan2006firstcities}
O'Sullivan, A.: The First Cities, chap.~3, pp. 40--54.
\newblock John Wiley \& Sons, Ltd (2006).
\newblock \doi{https://doi.org/10.1002/9780470996225.ch3}

\bibitem{ottaviano2004agglomeration}
Ottaviano, G., Thisse, J.F.: {Chapter 58 - Agglomeration and Economic
  Geography}.
\newblock In: J.V. Henderson, J.F. Thisse (eds.) Cities and Geography,
  \emph{Handbook of Regional and Urban Economics}, vol.~4, pp. 2563--2608.
  Elsevier (2004).
\newblock \doi{https://doi.org/10.1016/S1574-0080(04)80015-4}.
\newblock
  \urlprefix\url{https://www.sciencedirect.com/science/article/pii/S1574008004800154}

\bibitem{Patterson2015STDs}
Patterson-Lomba, O., Goldstein, E., G{\'o}mez-Li{\'e}vano, A., Castillo-Chavez,
  C., Towers, S.: Per capita incidence of sexually transmitted infections
  increases systematically with urban population size: a cross-sectional study.
\newblock Sexually Transmitted Infections pp. 1--5 (2015)

\bibitem{paulus2006evolutionary}
Paulus, F., Pumain, D., Vacchiani-Marcuzzo, C., Lobo, J.: An evolutionary
  theory for interpreting urban scaling laws.
\newblock Cybergeo: Revue europ{\'e}enne de g{\'e}ographie/European journal of
  geography \textbf{{}}(343), 20 (2006)

\bibitem{pressman1971crime}
Pressman, I., Carol, A.: Crime as a diseconomy of scale.
\newblock Review of Social Economy \textbf{29}(2), 227--236 (1971)

\bibitem{pumain2018evolutionary}
Pumain, D.: An evolutionary theory of urban systems.
\newblock In: {Rozenblat C., Pumain D., Velasquez E. International and
  Transnational Perspectives on Urban Systems}, pp. 3--18. Springer (2018)

\bibitem{ReichDavid2018Wwaa}
Reich, D.: Who we are and how we got here : ancient DNA and the new science of
  the human past, first edition edn.
\newblock Pantheon Books, New York (2018)

\bibitem{ribeiro2021mathematical}
Ribeiro, F.L., Rybski, D.: Mathematical models to explain the origin of urban
  scaling laws: a synthetic review.
\newblock arXiv preprint arXiv:2111.08365  (2021)

\bibitem{ribeiro2018unveiling}
Ribeiro, H.V., Hanley, Q.S., Lewis, D.: Unveiling relationships between crime
  and property in england and wales via density scale-adjusted metrics and
  network tools.
\newblock PLoS One \textbf{13}(2), e0192931 (2018)

\bibitem{richerson2008migration}
Richerson, P.J., Boyd, R.: Migration: An engine for social change.
\newblock Nature \textbf{456}(7224), 877--877 (2008)

\bibitem{rotolo2006population}
Rotolo, T., Tittle, C.R.: Population size, change, and crime in us cities.
\newblock Journal of Quantitative Criminology \textbf{22}(4), 341--367 (2006)

\bibitem{sarkar2020evidence}
Sarkar, S., Arcaute, E., Hatna, E., Alizadeh, T., Searle, G., Batty, M.:
  Evidence for localization and urbanization economies in urban scaling.
\newblock {Royal Society Open Science} \textbf{7}(3), 191638 (2020)

\bibitem{schroeder1991fractals}
Schroeder, M.: Fractals, chaos, power laws: Minutes from an infinite paradise.
\newblock Dover (1991)

\bibitem{seiyama1986review}
Seiyama, K.: A review of mathematical models of formal organizations—why and
  how they failed—.
\newblock Journal of Mathematical Sociology \textbf{12}(1), 71--96 (1986)

\bibitem{shalizi2011scaling}
Shalizi, C.R.: Scaling and hierarchy in urban economies.
\newblock arXiv preprint arXiv:1102.4101  (2011)

\bibitem{shennandemography2001}
Shennan, S.: Demography and cultural innovation: a model and its implications
  for the emergence of modern human culture.
\newblock Cambridge archaeological journal \textbf{11}(1), 5 (2001)

\bibitem{simon1955behavioral}
Simon, H.A.: A behavioral model of rational choice.
\newblock {The Quarterly Journal of Economics} pp. 99--118 (1955)

\bibitem{Sornette2006}
Sornette, D.: {Critical Phenomena in Natural Sciences: Chaos, Fractals,
  Selforganization and Disorder: Concepts and Tools}, 2nd edn.
\newblock Springer Series in Synergetics. Springer, Heidelberg (2006)

\bibitem{starrett1978market}
Starrett, D.: Market allocations of location choice in a model with free
  mobility.
\newblock Journal of economic theory \textbf{17}(1), 21--37 (1978)

\bibitem{stiglitz2018modern}
Stiglitz, J.E.: Where modern macroeconomics went wrong.
\newblock Oxford Review of Economic Policy \textbf{34}(1-2), 70--106 (2018)

\bibitem{strumsky2005metropolitan}
Strumsky, D., Lobo, J., Fleming, L.: Metropolitan patenting, inventor
  agglomeration and social networks: A tale of two effects.
\newblock SFI Working Paper 2005-02-004, Santa Fe Institute, Santa Fe, NM
  (2005)

\bibitem{Sveikauskas1975}
Sveikauskas, L.: {The Productivity of Cities}.
\newblock The Quarterly Journal of Economics \textbf{89}(3), 393--413 (1975).
\newblock \urlprefix\url{http://www.jstor.org/stable/1885259}

\bibitem{thisse2014new}
Thisse, J.F.: The new science of cities by michael batty: The opinion of an
  economist.
\newblock Journal of Economic Literature \textbf{52}(3), 805--19 (2014)

\bibitem{weitzman1998recombinant}
Weitzman, M.L.: Recombinant growth.
\newblock The Quarterly Journal of Economics \textbf{113}(2), 331--360 (1998)

\bibitem{west1997general}
West, G.B., Brown, J.H., Enquist, B.J.: A general model for the origin of
  allometric scaling laws in biology.
\newblock Science \textbf{276}(5309), 122--126 (1997)

\bibitem{wirth1938urbanism}
Wirth, L.: Urbanism as a way of life.
\newblock American journal of sociology \textbf{44}(1), 1--24 (1938)

\bibitem{YounEtAl2016Universality}
Youn, H., Bettencourt, L.M.A., Lobo, J., Strumsky, D., Samaniego, H., West,
  G.B.: Scaling and universality in urban economic diversification.
\newblock Journal of The Royal Society Interface \textbf{13}(114) (2016).
\newblock \doi{10.1098/rsif.2015.0937}

\bibitem{zumel2020innovation}
Zumel~Dumlao, J.M., Lei, J., Nwosu, E., Oon, L.Y., Wong, T.L.J., Rising, J.,
  Anttila-Hughes, J.: Innovation dynamics of cultural production: Evidence in
  rap lyrics.
\newblock Tech. rep., University of San Francisco, San Francisco, CA (2020).
\newblock \urlprefix\url{https://repository.usfca.edu/thes/1315}

\end{thebibliography}
\bibliographystyle{spmpsci}
\end{document}